\documentclass[twocolumn, tighten]{aastex631}
\usepackage{amssymb}
\usepackage{amsmath}
\usepackage{gensymb}
\usepackage{subfigure}
\usepackage{natbib}
 \hypersetup{
 linkcolor=blue,
 citecolor=blue,
 filecolor=blue,
 urlcolor=blue
 }

\newcommand{\SE}{\texttt{SExtractor}}
\newcommand{\fluxauto}{\texttt{FLUX\textunderscore AUTO}}
\newcommand{\fluxrad}{\texttt{FLUX\textunderscore RADIUS}}
\newcommand{\starflag}{\texttt{star}}
\newcommand{\detcontam}{\texttt{det\textunderscore contam}}
\newcommand{\eazypy}{\texttt{eazy-py}}
\newcommand{\EAZY}{\texttt{EAZY}}
\newcommand{\usephot}{\texttt{use\_phot}}

\shorttitle{UDS catalog}
\shortauthors{Zaidi}

\graphicspath{{./}}

\begin{document}

\title{The FENIKS Survey: Multi-wavelength Photometric Catalog in the UDS Field, and Catalogs of Photometric Redshifts and Stellar Population Properties

}

\author[0000-0002-1163-7790]{Kumail Zaidi}
\affiliation{Department of Physics \& Astronomy, Tufts University, 574 Boston Avenue, Medford, MA 02155, USA}

\author[0000-0001-9002-3502]{Danilo Marchesini}
\affiliation{Department of Physics \& Astronomy, Tufts University, 574 Boston Avenue, Medford, MA 02155, USA}

\author[0000-0001-7503-8482]{Casey Papovich}
\affiliation{George P. and Cynthia Woods Mitchell Institute for Fundamental Physics and Astronomy, Texas A\&M University, College Station, TX 78743, USA}
\affiliation{Department of Physics \& Astronomy, Texas A\&M University, 4242 TAMU, College Station, TX 78743, USA}

\author[0000-0002-0243-6575]{Jacqueline Antwi-Danso}
\affiliation{George P. and Cynthia Woods Mitchell Institute for Fundamental Physics and Astronomy, Texas A\&M University, College Station, TX 78743, USA}
\affiliation{Department of Physics \& Astronomy, Texas A\&M University, 4242 TAMU, College Station, TX 78743, USA}
\affiliation{David A. Dunlap Department of Astronomy \& Astrophysics, University of Toronto, 50 St George St, Toronto, ON M5S 3H4, Canada}

\author[0000-0001-6342-9662]{Mario Nonino}
\affiliation{INAF-Osservatorio Astronomico di Trieste, Via Bazzoni 2, 34124 Trieste, Italy}

\author[0000-0002-8053-8040]{Marianna Annunziatella}
\affiliation{Centro de Astrobiología (CAB), CSIC-INTA, Carretera de Ajalvir km 4, Torrejón de Ardoz, E-28850, Madrid, Spain}
\affiliation{INAF-Osservatorio Astronomico di Capodimonte, Via Moiariello 16, I-80131 Napoli, Italy}

\author[0000-0003-2680-005X]{Gabriel Brammer}
\affiliation{Cosmic Dawn Center (DAWN), Niels Bohr Institute, University of Copenhagen, Jagtvej 128, København N, DK-2200, Denmark}

\author[0000-0001-6941-7662]{James Esdaile}
\affiliation{Centre for Astrophysics and Supercomputing, Swinburne University of Technology, Melbourne, VIC 3122, Australia}
\affiliation{ARC Centre for Excellence in All-Sky Astrophysics in 3D (ASTRO 3D), Australia}

\author[0000-0002-3254-9044]{Karl Glazebrook}
\affiliation{ARC Centre of Excellence for All Sky Astrophysics in 3 Dimensions (ASTRO 3D), Australia}
\affiliation{Centre for Astrophysics and Supercomputing, Swinburne University of Technology, Melbourne, VIC 3122, Australia}

\author[0000-0001-9298-3523]{Kartheik Iyer}
\affiliation{David A. Dunlap Department of Astronomy \& Astrophysics, University of Toronto, 50 St George St, Toronto, ON M5S 3H4, Canada}
\affiliation{Columbia Astrophysics Laboratory, Columbia University, 550 West 120th Street, New York, NY 10027, USA}

\author[0000-0002-2057-5376]{Ivo Labbé}
\affiliation{Centre for Astrophysics and Supercomputing, Swinburne University of Technology, Melbourne, VIC 3122, Australia}

\author[0000-0002-7248-1566]{Z. Cemile Marsan}
\affiliation{Department of Physics \& Astronomy, York University, 4700 Keele Street Toronto, Ontario, M3J 1P3, Canada}

\author[0000-0002-9330-9108]{Adam Muzzin}
\affiliation{Department of Physics \& Astronomy, York University, 4700 Keele Street Toronto, Ontario, M3J 1P3, Canada}

\author[0000-0002-6047-1010]{David A. Wake}
\affiliation{Department of Physics \& Astronomy, University of North Carolina Asheville, Asheville, NC 28804, USA}

\correspondingauthor{Kumail Zaidi}
\email{kumail.zaidi@tufts.edu}

\submitjournal {ApJ}
\begin{abstract}
We present the construction of a deep multi-wavelength PSF-matched photometric catalog in the UDS field following the final UKIDSS UDS release. The catalog includes photometry in 24 filters, from the MegaCam-$uS$ 0.38$\mu$m band to the {\it Spitzer}-IRAC 8$\mu$m band, over $\sim$ $0.9$ deg$^{2}$ and with a 5$\sigma$ depth of $25.3$ AB in the $K$-band detection image. The catalog,  containing $\approx 188, 564$ $(136, 235)$ galaxies at $0.2<z<8.0$ with stellar mass $\log{(M_{*}/M_{\odot})}>8$ and $K$-band total magnitude $K<25.2$ $(24.3)$ AB, enables a range of extragalactic studies. We also provide photometric redshifts, corresponding redshift probability distributions, and rest-frame absolute magnitudes and colors derived using the template-fitting code \eazypy. Photometric redshift errors are less than $3-4\%$ at $z<4$ across the full brightness range in $K$-band and stellar mass range $8<\log{(M_{*}/M_{\odot})}<12$. Stellar population properties (e.g., stellar mass, star-formation rate, dust extinction) are derived from the modeling of the spectral energy distributions (SEDs) using the codes \texttt{FAST} and Dense Basis. 
\end{abstract}


\section{Introduction}\label{sec:intro}

Over the last 15 years, the construction of near-infrared (NIR) detected multi-wavelength photometric survey catalogs has become progressively more sophisticated (MUSYC, \citealt{2007AJ....134.1103Q}; NMBS, \citealt{2011ApJ...735...86W}; UVISTA-DR1, \citealt{2013ApJS..206....8M}; ZFOURGE, \citealt{2016ApJ...830...51S}; COSMOS2015, \citealt{Laigle_2016}; COSMOS2020, \citealt{2022ApJS..258...11W}).

The high-quality, point-spread function (PSF) matched multi-wavelength photometry is essential for any investigation of galaxy evolution. Besides allowing for large photometric statistical analysis, they enable the identification of robust targets for detailed follow-up studies, e.g., spectroscopy for confirmation and stellar continuum and emission line studies, longer wavelength investigations (e.g., with ALMA), or space-based imaging and spectroscopy.

This has become increasingly important in the era of JWST, as the previously unprobed regions of parameter space (e.g., low-mass galaxies and objects in the epoch of re-ionization) by ground-based facilities can now be explored. For instance, rest-frame optical spectroscopic studies of massive dusty star-forming galaxies at $z>4$ are now possible with JWST. To facilitate such endeavors, here we present a $K$-band detected, PSF-matched multi-wavelength (from the $u$-band to the mid-IR) photometric catalog for the F2 Extragalactic NearIR K-Split Survey (FENIKS). The FENIKS survey is currently being carried out using two medium-band filters, K$_{blue}$ ($\lambda_{eff}$ = 2.06$\mu$m) and K$_{red}$ ($\lambda_{eff}$ = 2.31$\mu$m) (see \citealp{2021AJ....162..225E} for more details) recently added to the \texttt{FLAMINGOS-2 (F2)} instrument (\citealp{2008SPIE.7014E..0VE, 2012SPIE.8446E..0IE, 2012AAS...21941307G}) atop the 8.1-meter Gemini South Telescope in Chile. The medium K$_{blue}$/K$_{red}$ filters, which straddle the K-band, increase the photometric redshift accuracy with better sampling of either side of the Balmer/4000\AA\ breaks of the $z > 4$ galaxies. 

The footprint of the catalog and the primary data comes from the UKIDSS UDS data release 11 (UDS DR11, hereafter) for which other catalogs already exist (Almaini et al. in prep.; Hartley et al. in prep.)\footnote{\url{https://www.nottingham.ac.uk/~ppzoa/DR11/}} The UKIDSS project is defined in \cite{2007MNRAS.379.1599L}. Further details on the UDS can be found in Almaini et al. (in prep). UKIDSS uses the UKIRT Wide Field Camera (WFCAM; \cite{2007A&A...467..777C}). The photometric system is described in \cite{2006MNRAS.367..454H}, and the calibration is described in \cite{2009MNRAS.394..675H}. The pipeline processing and science archive are described in Irwin et al (in prep) and \cite{2008MNRAS.384..637H}. Besides the data in the broad-band filters of UDS DR11 ($J$, $H$, $K$) and the FENIKS medium bands, we also included ancillary data from other surveys: CLAUDS (MegaCam $uS$-band; \citealt{2019MNRAS.489.5202S}), SXDS ($B$, $V$, $R$, $i$, $z$; \citealt{2008ASPC..399..131F}), HSC ($g$, $r$, $i$, $z$, $y$, NB0816, NB0921; \citealt{2019PASJ...71..114A}), VIDEO ($H$, $K_s$, $Y$, $z$; \citealt{2013MNRAS.428.1281J}), {\it Spitzer}-IRAC (ch1-4; \citealt{2012PASP..124..714M, 2021MNRAS.501..892L}), totaling data in 26 bands over a surveyed area of $\sim$ 0.9 deg$^2$. Our catalog, containing about half the area but $\sim$0.7(0.3)~mag deeper than the UltraVISTA $K$-band `deep'(`ultra-deep) regions, is nicely complementary to the COSMOS2020 catalog \citep{2022ApJS..258...11W}. This makes the FENIKS catalog the deepest wide-field ground-based NIR dataset and importantly, it probes a different line of sight, vital in reducing the uncertainties due to cosmic variance in measured quantities (e.g., number densities).

The photometric catalog, along with the associated catalogs of photometric redshifts and stellar population properties, are made publicly available\footnote{The Version 1 (v1) release of the FENIKS UDS catalogs containing photometric catalog, the redshift catalog and the catalogs of stellar population properties as described in this paper can be found at \url{https://www.zaidikumail.com/feniks-uds-catalogs} and through Zenodo at \url{10.5281/zenodo.11002299}}, enabling scientific investigations over a wide range in redshift ($0.2<z<8.0$, i.e., covering about 95\% of cosmic history). The combination of very deep NIR imaging over a wide area allows for the study of both distant faint/low-mass galaxies, as well as the population of rare distant bright/massive galaxies. The ultra-deep $K$-band photometry in combination with high-quality photometric redshifts enables complete studies of galaxies out to $z\approx 5$, as the selection in the rest-frame optical reduces biases against evolved and/or dusty star-forming galaxies.

\begin{figure*}[htb!]
\includegraphics[width=\textwidth]{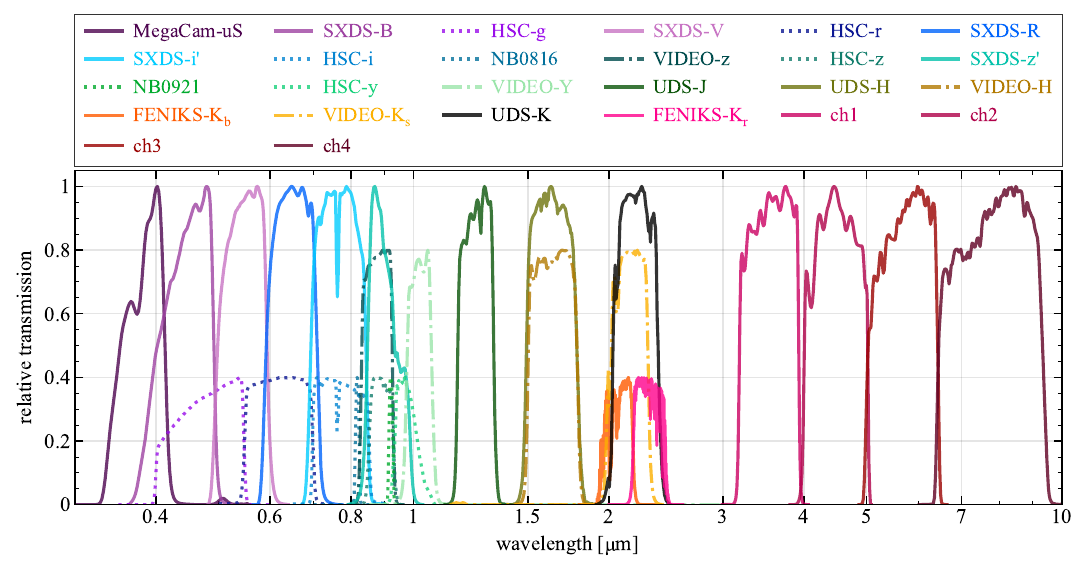}
\caption{Filter transmission curves of different bands. For clarity, some filters are normalized to have a maximum relative transmission less than 1: FENIKS (solid) $K_{blue}$ and $K_{red}$ peak at 0.4; HSC (dotted) $g$, $r$, $i$, $z$, $y$, NB0816, and NB0921 have peaks at 0.4; VIDEO (dash-dotted) $z$, $Y$, $H$, and $K_{s}$ peak at 0.8. All the other bands (solid) peak at 1. The detection UDS $K$-band  (see \S \ref{subsec:sex} for details) is shown in solid black.}
\label{fig:tcurves}
\end{figure*}

The paper is structured as follows: in \S \ref{sec:data}, we describe the different surveys from which the multi-wavelength data have been combined to build the catalog; in \S \ref{sec:catcons}, we present the procedure for the construction of the catalog in detail, as well as define the included quantities/parameters in the catalog and how they can be used to make a typical selection of galaxies; \S\ref{sec:z-stellar} presents the modeling of the spectral energy distributions (SEDs) to derive photometric redshifts and stellar population properties; we conclude with a summary in \S \ref{sec:summary}. 

Throughout this paper, we assume a $\Lambda$CDM cosmology with \textit{H$_{0}$} = 70 km s$^{-1}$ Mpc$^{-1}$, $\Omega_{m, 0}$ = 0.3, and $\Omega_{\Lambda, 0}$ = 0.7. The magnitudes and colors are presented in the AB magnitude system \citep{1983ApJ...266..713O}. The stellar population parameters are derived assuming the Chabrier initial mass function (IMF) \citep{2003PASP..115..763C}.

\section{Data and Overview}\label{sec:data}

\begin{figure*}[hbt!]
\includegraphics[width=\textwidth]{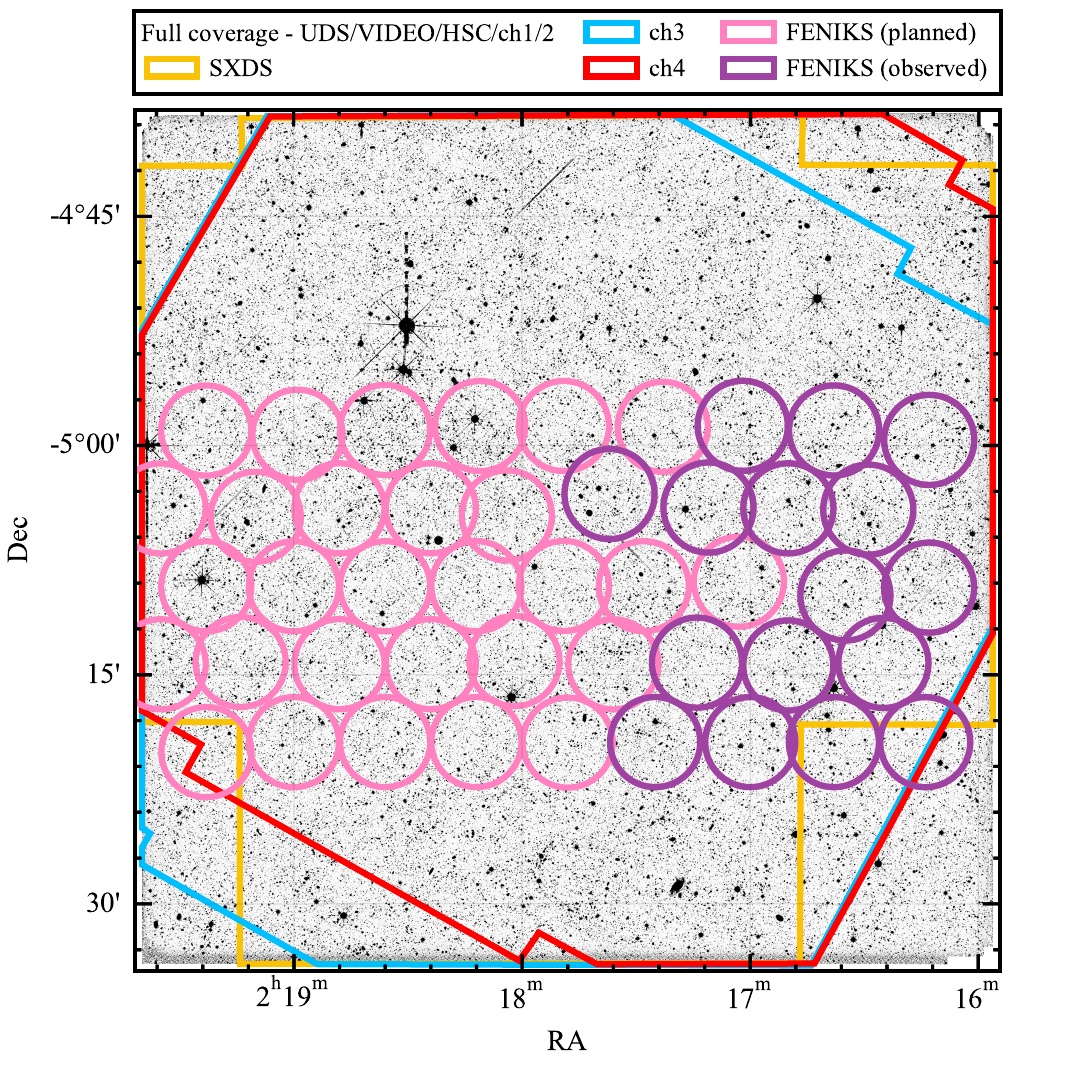}
\caption{The $K$-band (detection band) image overlaid with the footprints of different filters or surveys. The UDS, VIDEO, HSC, and IRAC ch1/2 bands (16 out of 26 bands utilized in the catalog) have full coverage across the shown $K$-band image. SXDS coverage is shown in yellow, IRAC ch3 in blue, IRAC ch4 in red, observed FENIKS $K_{blue}$/$K_{red}$ in purple, and planned FENIKS $K_{blue}$/$K_{red}$ in pink.}
\label{fig:coverage}
\end{figure*}

Our catalog is based on the final release (UDS DR11)\footnote{\url{https://www.nottingham.ac.uk/astronomy/UDS/data/dr11.html}} of the $J$, $H$, and $K$ data over the full UDS field which covers an area of $\sim$0.9 deg$^2$.  Along with these primary data, ancillary data have been combined from various deep surveys to produce a PSF-matched photometric catalog of 26 bands, whose transmission curves are shown in Fig. \ref{fig:tcurves}, the footprint shown in Fig.~\ref{fig:coverage}, and the metadata is shown in Table~\ref{tab:filters}.

In the optical, we employ the deep imaging available in $B$, $V$, $R$, $i$\textquotesingle, $z$\textquotesingle\ from the Subaru‒XMM-Newton Deep Survey (SXDS), which covers the UDS field with five SuprimeCam pointings from the 8.2m Subaru Telescope \citep{2008ASPC..399..131F}. We utilize the version of these mosaics generated by C. Simpson (Gemini) to complement the UDS DR11 release which benefits from a considerably better zero-point (ZP, hereafter) scaling and background subtraction compared to the initial release\footnote{\url{https://www.nottingham.ac.uk/~ppzdm/research.html}}. We target the shortest wavelength in our catalog using the available data in the $uS$ filter obtained with MegaCam on the Canada-France Hawaii Telescope (CFHT) as part of the CLAUDS survey \citep{2019MNRAS.489.5202S}.

Additionally, we took advantage of the overlap with the public data release 2 (PDR2) of the Hyper Suprime-Cam (HSC) Subaru Strategic Program (SSP) \citep{2019PASJ...71..114A}, also conducted by the Subaru Telescope. The PDR2 provides multi-band data in a large area of over 300~deg$^2$ including deep/ultra-deep and wide regions out of which the tract number 8523 overlaps with our UDS field in $g$, $r$, $i$, $z$, $Y$, $NB0921$, and $NB0816$ bands, all of which we included in our catalog. We noticed that the astrometry of the PDR2 was offset from UDS DR11 in both the RA and DEC direction, with median offsets of $\sim$ 0.1\arcsec. Furthermore, these offsets were not uniform across the image. To remedy this, we re-mapped the astrometry of the relevant part of PDR2 images using the centroids of stars relative to the corresponding stars in UDS DR11, resulting in residual offsets of the order of milli-arcseconds.

Available data from the data release 4 (DR4) of the VISTA Deep Extragalactic Observations (VIDEO; \citealt{2013MNRAS.428.1281J}) in the NIR filters $Z$, $Y$, $H$, $K_s$ were also included. VIDEO is a deep NIR imaging survey covering $\sim$12~deg$^2$ in three fields XMM1, XMM2, and XMM3 out of which XMM1 fully encapsulates the UDS DR11 pointing.

Finally, in the mid-IR, we took advantage of the data from the SERVS \citep{2012PASP..124..714M} and DeepDrill \citep{2021MNRAS.501..892L} surveys using the \textit{Spitzer Space Telescope} Infrared Array Camera (IRAC) in ch1 and ch2. For IRAC ch3 and ch4, we used the imaging data from the \dataset[SpUDS survey]{https://www.ipac.caltech.edu/doi/irsa/10.26131/IRSA403}.

Due to the delays in the FENIKS data collection in the medium $K_{blue}$ and $K_{red}$ bands at the Gemini Telescope, we have not incorporated the FENIKS medium-band photometry in our catalog yet. As soon as the data collection is complete, we will release an updated version of the publicly released catalogs including the FENIKS photometry along with the updated photometric redshifts and stellar population properties.

\section{Catalog construction} \label{sec:catcons}

\subsection{PSF matching of the optical-NIR bands}\label{subsec:PSFmatching}

The optical-NIR images in the 20 filters between MegaCam-$uS$ (0.38$\mu$m) and the UDS $K$ (2.1$\mu$m) bands suffer from atmospheric seeing resulting in varying image resolutions that manifest in inhomogeneous point spread functions (PSFs). As a result, aperture photometry cannot be performed consistently across the bands as a fixed aperture size would not necessarily correspond to the same physical region of the source. To ensure consistent aperture photometry, we need to homogenize the PSFs among all the optical-NIR bands. To do so, we first determined the lowest resolution band to which all the other bands should be PSF-matched. As a side note, the IRAC images are heavily blended and have much broader PSFs than any of the optical-NIR bands, and therefore they require a different approach to obtain consistent aperture photometry which we describe in $\S$\ref{subsec:irac}.

In practice, the median full-width-at-half-maximum (FWHM) of each band is determined by fitting a 2D Gaussian or Moffat profile on high signal-to-noise ratio stars and taking the mean or median of the resulting distribution. Then, the band with the largest FWHM is considered to be the worst resolution band. However, any variability in the PSF shapes of the different bands implies that the FWHMs resulting from simple Moffat or Gaussian fits are sometimes not appropriate to reliably identify the band with the worst image quality, especially if multiple bands have very similar FWHMs. Therefore, we adopted a much simpler approach assuming no functional form for the PSF shape. Specifically, for each band, we extracted the median growth curve using the normalized circular star stamps with a diameter equal to 32 pixels (8.6\arcsec), large enough to contain all of the stars' fluxes, and the corresponding median 50\% and 75\% encircled light radii, r$_{50}$, and r$_{75}$, respectively. The band with the worst resolution is MegaCam-$uS$ (r$_{50}$ $\sim$ 0.63\arcsec\ ; FWHM $\sim 1.00$\arcsec) which is slightly worse than the second worst resolution band, VIDEO-Y (r$_{50}$ $\sim$ 0.60\arcsec\ ; FWHM $\sim$ 0.87\arcsec). To avoid the extra loss of signal we would have had if all optical-NIR bands were PSF-matched to MegaCam-$uS$ instead of VIDEO-Y, we decided to PSF match all the optical-NIR bands (except MegaCam-$uS$) to VIDEO-Y.
\startlongtable
\begin{deluxetable*}{ccccccc}
\tablecolumns{6}
\tablewidth{\textwidth} 
\tablecaption{Metadata of the filters used in the catalog \label{tab:filters}}
\tablehead{
{Instrument/Telescope/Survey}& 
{Filter}&
{$\lambda_{eff}^{a}$ [\AA]}& 
{A$^{b}$ [mag]}&
{ZP offset$^{c}$}&
{3$\sigma$/5$\sigma$ depth$^{d}$ [AB]}&
{\textit{r$_{50}$$^{e}$} [\arcsec]}
}
\startdata
{MegaCam/CFHT/CLAUDS}& {uS}&{3828}&{0.085}&{1.068}&{27.62 / 27.07}&{0.625}\\
\hline
{Suprime-Cam/}&\textit{B}&{4448}&{0.075}&{1.011}&{28.14 / 27.58}&{0.554}\\
{Subaru XMM-Newton}&\textit{V}&{5470}&{0.058}&{0.997}&{27.86 / 27.31}&{0.540}\\
{survey (SXDS)}&\textit{R}&{6505}&{0.045}&{1.004}&{27.60 / 27.05}&{0.547}\\
{}&\textit{i\textquotesingle}&{7671}&{0.035}&{1.023}&{27.41 / 26.86}&{0.560}\\
{}&\textit{z\textquotesingle}&{9028}&{0.027}&{1.049}&{26.59 / 26.03}&{0.538}\\
\hline
{HSC/}&\textit{g}&{4798}&{0.069}&{0.989}&{27.70 / 27.14}&{0.477}\\
{Subaru/}&\textit{r}&{6218}&{0.048}&{0.956}&{27.25 / 26.69}&{0.508}\\
{SSP}&\textit{i}&{7727}&{0.034}&{0.986}&{27.08 / 26.52}&{0.443}\\
{(PDR2)}&\textit{NB0816}&{8177}&{0.032}&{1.001}&{26.44 / 25.89}&{0.412}\\
{}&\textit{z}&{8908}&{0.027}&{0.941}&{26.40 / 25.85}&{0.472}\\
{}&\textit{NB0921}&{9213}&{0.026}&{0.966}&{26.18 / 25.62}&{0.536}\\
{}&\textit{y}&{9775}&{0.024}&{0.930}&{25.62 / 25.06}&{0.447}\\
\hline
{VISTA/}&\textit{z}&{8787}&{0.028}&{0.937}&{26.02 / 25.46}&{0.592}\\
{VIDEO}&\textit{Y}&{10217}&{0.022}&{0.951}&{25.49 / 24.94}&{0.593}\\
{(DR4)}&\textit{H}&{16433}&{0.010}&{1.038}&{24.39 / 23.84}&{0.531}\\
{}&\textit{Ks}&{21503}&{0.007}&{1.010}&{24.32 / 23.76}&{0.534}\\
\hline
{UKIRT/}&\textit{J}&{12502}&{0.015}&{1.002}&{26.15 / 25.60}&{0.531}\\
{UDS}&\textit{H}&{16360}&{0.010}&{1.002}&{25.53 / 24.98}&{0.539}\\
{(DR11)}&\textit{K}&{22060}&{0.006}&{1.000}&{25.90 / 25.35}&{0.503}\\
\hline
{IRAC/\textit{Spitzer}}&{ch1}&{35200}&{0.004}&{1.002}&{24.74 / 24.19}&{-}\\
{(SERVS/DeepDrill)}&{ch2}&{44600}&{0.003}&{1.017}&{24.31 / 23.75}&{-}\\
\hline
{IRAC/\textit{Spitzer}}&{ch3}&{56600}&{0.003}&{0.870}&{22.72 / 22.17}&{-}\\
{(SpUDS)}&{ch4}&{76800}&{0.003}&{0.908}&{22.62 / 22.07}&{-}\\
\enddata
\tablecomments{\\
$^{a}$Effective wavelength; $^{b}$Galactic dust extinction from \cite{2011ApJ...737..103S} calculated using the online tool at \url{https://irsa.ipac.caltech.edu/applications/DUST/}; $^{c}$ZP offsets calculated using \eazypy\ with respect to the $K$-band - this multiplicative factor was applied to the photometry to derive photometric redshifts and stellar population properties; $^{d}$3$\sigma$ and 5$\sigma$ depths of the PSF-matched images scaled to total using the final PSF-matched growth curve, except for MegaCam-\textit{uS}. For MegaCam-\textit{uS}, the depths were calculated using the non-PSF-matched image and scaled to the total using the corresponding growth curve; $^{e}$median 50\%\ encircled light radius derived using the stars from the non-PSF-matched images to assess image quality.
}
\end{deluxetable*}
In \S\ref{subsec:aperphot}, we describe how we differently included PSF-matched aperture photometry for the MegaCam-$uS$ band.

The FWHMs of stars are found to vary between +/-2\% and +/-4\% across the field, depending on the band, which is expected as the image footprint is large enough ($\sim$0.9 deg$^2$). Therefore, instead of having one single model PSF for the whole image, we subdivided our images into grids, determined the model PSFs for each patch, and then PSF matched the images patch by patch. To determine model PSFs, both in the low-resolution image (VIDEO-$Y$) and high-resolution images, we utilized unsaturated, high S/N stars in the local area, enough in number to sample the local PSF appropriately. The density of the PSF grid was based on the number of total stars selected in each high-resolution image. Once the PSF grid for a high-resolution band and the corresponding grid for VIDEO-$Y$ is developed, we fit the Fourier transform of the model PSFs with a series of Gaussian-weighted Hermite Polynomials as presented in \cite{2014ApJS..214...24S} and \cite{2018ApJS..235...14S} to obtain the corresponding convolution kernel grid. The convolution kernels were then used to PSF match the whole image patch by patch, and then stitched together using the code TOPH\footnote{\url{https://github.com/jacqdanso/TOPH}}.

\begin{figure*}[hbt!]
\includegraphics[width=\textwidth]{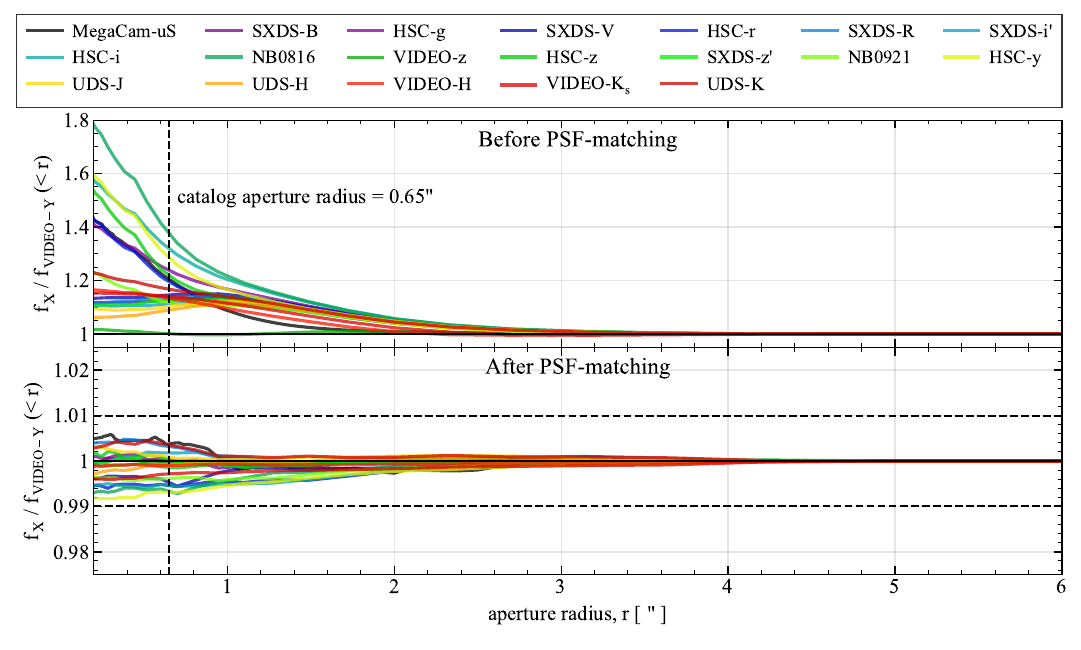}
\caption{Flux ratios (f$_{\,X}$ / f$_{\,VIDEO-Y}$) of the median growth curves of the stars in the individual bands, X, and VIDEO-Y before ({\it top panel}) and after PSF matching ({\it bottom panel}), except for MegaCam-$uS$. The PSF-matching for MegaCam-$uS$ was treated differently  in which UDS-$K$ was PSF-matched to MegaCam-$uS$ (see \S \ref{subsec:aperphot}), therefore, we plot (f$_{\,UDS-K}$ / f$_{\,MegaCam-uS}$). These median growth curves are derived using the same cleaned star stamps utilized in the PSF matching. The vertical dashed line in both panels  represents the catalog aperture radius (color aperture) of 0.65\arcsec (aperture diameter of 1.3\arcsec). The median growth curves of the PSF-matched bands agree within 1\% at all radii, as shown in the bottom panel.}
\label{fig:gcurves}
\end{figure*}

Figure~\ref{fig:gcurves} shows the ratio of the high-resolution image median growth curves to the VIDEO-$Y$'s median growth curve before (top) and after (bottom) PSF matching. The median growth curves were derived from the same cleaned star stamps utilized in the PSF matching. After PSF matching, the median growth curves of all bands agree very well, within 1\%, at all radii.

\subsection{K-band Detection}\label{subsec:sex}
The UDS DR11 $K$-band, with a total 5-sigma depth of 25.3 AB, was used as the source detection band. We used \SE\ \citep{1996A&AS..117..393B} to perform the detection using the sky-subtracted UDS $K$-band image along with the associated weight map. A series of tests were performed to fine-tune the critical \SE\ detection parameters, namely \texttt{DETECT\_MINAREA}, \texttt{DETECT\_THRESH}, and \texttt{DEBLEND\_MINCONT}. 

The best combination of detection parameters was identified by inspecting the resulting segmentation map and the detected sources. Firstly, we ensured that the object centroids were located inside the pixel with peak flux value, and that resolved close pairs were not detected as single objects. Furthermore, occasionally a faint few pixels (usually noise) get assigned the same object ID as the brighter source in the vicinity, even if those faint pixels are not physically connected to the brighter source (quite common in \SE) i.e., a big `island' (brighter source) and a small `island' (a faint few pixels) get assigned the same object ID. This is regardless of whether the brighter source is a point source or not. We optimized the detection parameters by minimizing such occurrences, however, there still remained many such stray faint pixels assigned to the neighboring brighter sources. We set such pixels to a value of `0' in the segmentation map \textit{a posteriori}, meaning that those pixels are not assigned to any object. Finally, we arrived at the best set of detection parameters ensuring that no real object was undetected. The final set of chosen parameters is shown in Table~\ref{tab:SEparams}.

\begin{deluxetable}{ll}
\tablecolumns{2}
\tablewidth{0pt} 
\tablecaption{Adopted \SE\ detection and photometry parameters \label{tab:SEparams}}
\tablehead{
{Parameter} & {Value}
}
\startdata 
\texttt{ANALYSIS\_THRESH}& 0.9\\ 
\texttt{BACK\_FILTERSIZE}&{5}\\
\texttt{BACK\_SIZE}&{64}\\
\texttt{BACK\_TYPE }&{MANUAL}\\
\texttt{BACK\_VALUE}&{0}\\
\texttt{CLEAN}& Y\\
\texttt{CLEAN\_PARAM}& Y\\
\texttt{DETECT\_MINAREA}& 7\\
\texttt{DEBLEND\_MINCONT}&1x10$^{-7}$\\
\texttt{DEBLEND\_NTHRESH}& 32\\
\texttt{DETECT\_TYPE}& CCD\\
\texttt{FILTER}& Y\\
\texttt{FILTER\_NAME}& gauss\_3.0\_7x7.conv\\ 
\texttt{MASK\_TYPE}& CORRECT\\
\texttt{PHOT\_APERTURES}&2.235, 3.726, 4.844, 5.589, \\ {}&7.452,11.177, 14.903, 18.629\\
\texttt{PHOT\_AUTOPARAMS}&{2.5, 3.5}\\
\texttt{PIXEL\_SCALE}&{0.2684}\\
\texttt{SATUR\_LEVEL}&{150000}\\
\texttt{STARNNW\_NAME}&{default.nnw}\\
\texttt{THRESH\_TYPE}&RELATIVE\\
\texttt{WEIGHT\_TYPE}&{MAP\_WEIGHT,NONE}
\enddata
\end{deluxetable}

\subsection{Aperture photometry} \label{subsec:aperphot}

To obtain aperture photometry, we first determined the optimal aperture diameter which would maximize the signal-to-noise ratio (S/N). To do so, similar to the extraction of growth curves before PSF-matching above using bright and unsaturated stars, we extracted the growth curves after PSF-matching using the same diameter of 32 pixels (8.6\arcsec) used for normalization of the circular apertures. Then, using the resulting median growth curves in each band, we calculated the corresponding S/N as a function of aperture radius, r, as follows:
\begin{equation}
  S/N_{X} (r) = \frac{\tilde{g}_{X}(r)}{\sqrt{\tilde{g}_{X}(r) + 4{\pi}r^2}},
\end{equation}
where  $\tilde{g}_X$ denotes the median PSF-matched growth curve of any band (denoted by $X$), and S/N$_{X}$ is the corresponding S/N curve. The median of the aperture diameters corresponding to the peak S/N in all bands - D=1.3\arcsec\ - was adopted as the optimal (color) aperture diameter to be used to derive all color fluxes $f^{col}_X$ in each band $X$. We ran \SE\ in dual mode using the parameter values shown in Table~\ref{tab:SEparams} to detect sources and extract the photometry for the detected objects in the PSF-matched bands within the optimal color aperture diameter. 

It should be noted that as mentioned in \S\ref{subsec:PSFmatching}, we do not PSF-match MegaCam-$uS$ to VIDEO-Y to obtain its optimal aperture photometry like the rest of the optical-NIR bands. Instead, following \cite{2007AJ....134.1103Q}, we take advantage of the assumption that the $(\textit{u - K})$ color within D=1.3\arcsec at VIDEO-Y's resolution, $(u - K) \vert_{D=1.3^"}^{@Y-res}$, is the same as $(\textit{u - K})$ color within D=1.5\arcsec at MegaCam-$uS$'s resolution, $(u - K) \vert_{D=1.5^"}^{@u-res}$. This allows for easy extraction of the optimal color aperture photometry for the MegaCam-$uS$ without the need to PSF-match it to VIDEO-Y. In Figure \ref{fig:gcurves}, we show the growth curve of the $K$-band PSF-matched to MegaCam-$uS$ (black curve labelled `MegaCam-uS'). As investigated by \cite{2007AJ....134.1103Q}, this assumption works reliably for distant galaxies that our catalog is optimized for. This is expected because distant galaxies are barely resolved by ground-based imaging in the color aperture sizes considered here, hence any discrepancies in colors measured using different aperture sizes and/or different resolutions will be negligible \citep{2007AJ....134.1103Q}.

Furthermore, we measured the total flux for the objects in the $K$-band utilizing the \SE\ output FLUX\textunderscore AUTO - the flux within a flexible elliptical aperture \citep{1980ApJS...43..305K}. Although FLUX\textunderscore AUTO is a good approximation of the total flux, it is still an underestimation of it. We therefore applied an auto-to-total correction using the PSF-matched $K$-band growth curve, $\tilde{g}_{K}$, as shown in Eq.~\ref{eqn:ftotK}, where $r_{Kron}$ is the circularized Kron radius evaluated using the \SE's KRON\textunderscore RADIUS, A\textunderscore IMAGE, and B\textunderscore IMAGE, \textit{f$_{tot,K}$} is the total flux in the $K$-band, and \textit{f$_{AUTO,K}$} is the \SE's FLUX\textunderscore AUTO in $K$-band.

\begin{equation}\label{eqn:ftotK}
  f_{tot,K} = f_{AUTO,K} \times \frac{1}{\tilde{g}_{K} (r_{Kron})}
\end{equation}
This correction is smaller for large Kron radii, and in general, is on the order of $\sim 10\%$. We note that these corrections utilize $\tilde{g}_{K}$ which is derived from point sources, and therefore, might not be accurately applicable to extended faint sources outside of the AUTO aperture. However, any resulting bias will be sub-dominant in the presence of large flux uncertainties associated with extended faint sources \citep{2007AJ....134.1103Q}. We show the comparison of our total photometry with other catalogs in Appendix \ref{sec:mag_offsets}.

\subsection{Errors on photometry} \label{subsec:unc}

The errors on the aperture fluxes have contributions from background noise and Poisson noise. The error due to the background noise, $\sigma_{rms}$, depends on the extent of the correlation between the number of pixels, $N$, within the adopted apertures. If the pixels are completely uncorrelated, then $\sigma_{rms} \propto \sqrt{N}$. On the other hand, if the pixels are perfectly correlated, $\sigma_{rms} \propto N$. In reality, the correlation of the pixels is neither perfect nor absent but somewhere in between (\citealp{2003AJ....125.1107L, 2007AJ....134.1103Q, 2011ApJ...735...86W, 2016ApJ...830...51S,2016ApJS..224...28P}). On small scales, pixels are correlated due to interpolations performed on the local PSF. On large scales, there are various contributions, such as i) imperfect background subtraction, ii) the glow of the bright stars extending beyond the masked regions, and iii) faint, undetected sources.

Therefore, we determined the background noise for each band empirically by randomly placing 5000 empty apertures of diameter 1.3\arcsec\ (the color aperture diameter as described in \S \ref{subsec:aperphot}), avoiding the pixels occupied by the sources in the segmentation map.  Then, we fitted Gaussian functions to the resulting distributions and calculated the standard deviation, $\sigma_{rms,X}$ for each PSF-matched $X$-band. The 3-$\sigma$ and 5-$\sigma$ total depths listed in Table \ref{tab:filters} are calculated using this method. Finally, we calculate the total error budget for fluxes in the color aperture by adding the contribution from background noise and Poisson noise in quadrature. The error due to background dominates for the faint sources whereas the Poisson error dominates for the bright sources \citep{2011ApJ...735...86W}.

We similarly calculated the errors on the auto fluxes (fluxes enclosed by the Kron radii) in the $K$-band. As the Kron radius differs for each object, we model the background noise on auto fluxes, $\sigma_{auto, \space K}$ as a function of area, $A$, enclosed by the Kron radius as:
\begin{equation}
    \sigma_{auto, \space K} (A) = \alpha A^{\beta/2},
\end{equation}
where $\alpha$ and $\beta$  are constants that were empirically determined. We fit for $\alpha$ and $\beta$ by randomly placing 5000 empty apertures for each area based on the range of Kron radii available. We then calculated the error on the $K$-band auto fluxes in a similar way to how the errors in color apertures were calculated above by adding the contributions from background noise and the Possion error in quadrature. All fluxes and associated errors were then further scaled to an AB ZP of 25, and corrected for Milky Way extinction as listed in Table~\ref{tab:filters}.

\subsection{IRAC photometry} \label{subsec:irac}

The resolution of the IRAC images is significantly worse than the optical-NIR images. Therefore, we chose not to PSF match the optical-NIR bands to the IRAC PSF. That would have resulted in a serious blending of objects and an unnecessary loss of vital depth. However, we still needed to derive consistent photometry in color apertures from the IRAC images. To that end, we use an algorithm designed specifically to clean the blended photometry of the heavily confused images \citep{2006AAS...20913202L}.

The brighter sources impacting the blending in the IRAC images are also typically bright in the $K$-band. Therefore, the $K$-band image is used as the high-resolution template to deblend the IRAC photometry. In practice, we use the $K$-band image PSF-matched to the VIDEO-Y image and, hereafter, whenever the $K$-band image is mentioned, it is assumed that it is the PSF-matched to the VIDEO-Y, if not otherwise explicitly indicated. The segmentation map of the $K$-band image is used to identify the position and extent of the sources in the IRAC images. Then using bright, unsaturated, and isolated stars in the $K$ and IRAC images, convolution kernels are built to convolve the $K$-band image to the IRAC band resolution. A model is then fit to the $K$-band image convolved to the IRAC band resolution. Finally, for each object, instead of performing the photometry on the modeled image as the fit is never perfect, it is directly performed on the original IRAC image after the models of the neighbors have been subtracted out (see \citealt{2007ASPC..379..356W} for a more detailed description of this method). The adopted approach and the deblending code have been previously used in, e.g., \citealt{2010ApJ...725.1277M} (MUSYC), \citealt{2011ApJ...735...86W} (NMBS), \cite{2014ApJS..214...24S} (3D-HST), \citealt{2016ApJ...830...51S} (ZFOURGE), \citealt{2013ApJS..206....8M} (UltraVISTA DR1). 

We started by measuring the deblended IRAC fluxes in an aperture of diameter D=3\arcsec. The extracted deblended IRAC fluxes are then corrected to match the resolution of the template $K$-band (i.e., the resolution of the VIDEO-Y image). Finally, we obtained the IRAC fluxes in the color aperture, D=1.3\arcsec\, at the resolution of VIDEO-Y image to conform with the rest of the catalog by assuming that the $K$-IRAC color in D=1.3\arcsec\ is the same as $K$-IRAC color in D=3\arcsec. Therefore, the resulting IRAC fluxes in the color aperture homogeneously matching the color aperture fluxes of the optical-NIR bands, $f_{D=1.3^",\space IRAC|_{Y-res}}$, is calculated as follows:

\begin{equation}
  f_{D=1.3^",\space IRAC|_{Y-res}} = f_{D=3^",\space IRAC|_{Y-res}} \times \frac{f_{D=1.3^",\space K|_{Y-res}}}{f_{D=3^",\space K|_{Y-res}}},
\end{equation}
where, $f_{D=3^",\space IRAC|_{Y-res}}$ is the deblended IRAC flux within D=3\arcsec\ at the resolution of the VIDEO-Y band, $f_{D=1.3^",\space K|_{Y-res}}$ is the flux within D=1.3\arcsec\ in the $K$-band PSF-matched to the VIDEO-Y image, and $f_{D=3^",\space K|_{Y-res}}$ is the flux within D=3\arcsec\ measured on the $K$-band image PSF-matched to the VIDEO-Y image after the removal of the neighboring sources. We applied the same method to obtain catalog fluxes for all of the IRAC bands (ch1-ch4).

\subsection{Star classification} \label{subsec:stars}

\begin{figure*}[htb!]
\includegraphics[width=\textwidth]
{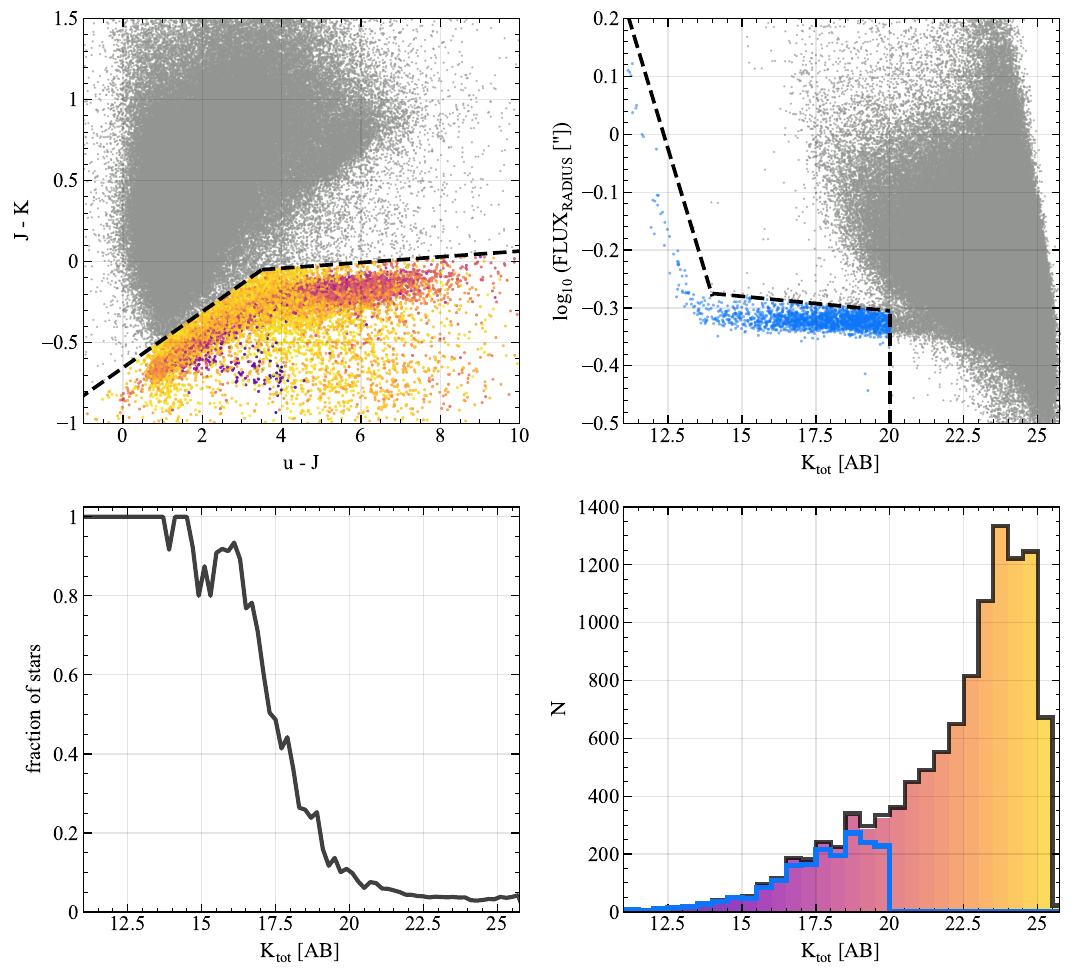}
\caption{Star selection using the $uJK$, and the $K_{tot}$ versus \SE's \fluxrad\ diagrams. \textbf{{\it Top-left:}} Observed $u-J$ versus  $J-K$ colors. The stars are selected as those objects falling below the plotted dashed lines (defined by Eq.~\ref{eqn:uJK_eqn}), and are shown as points with a yellow-purple color gradient indicating the $K$-band total magnitudes, as reflected in the histogram in the bottom-right panel. All other objects are shown in gray. \textbf{{\it Top-right:}} Additional star selection at the bright end ($K_{tot}<20$) using the $K_{tot}$ versus \SE's \fluxrad\ diagram. The objects falling below the dashed lines (defined by Eq.~\ref{eqn:fluxrad_v_K}) are classified as stars and shown in blue. Their $K_{tot}$ distribution is also shown in the bottom-right panel in blue. All other objects are again shown in gray. \textbf{{\it Bottom-left:}} The fraction of the stars in the catalog, selected using both the methods above (\starflag =1 in the catalog), as a function of $K_{tot}$ magnitude. The $K_{tot}$ distribution of this black curve is indicated in the bottom-right panel. \textbf{{\it Bottom-right:}} The $K_{tot}$ distribution of stars selected using the $uJK$ diagram (in purple-yellow), the $K_{tot}$ versus \SE's \fluxrad\ diagrams (in blue, applicable only for $K_{tot} < 20$ AB), and the combined final selection of stars (in black; \starflag = 1 in the catalog).}
\label{fig:stars}
\end{figure*}

First, we flagged our catalog sources as stars using their positions in the observed $u-J$ versus  $J-K$ color-color space ($uJK$ diagram). Stars segregate themselves from extended sources in such color-color diagrams which target similar combinations of wavelengths (see e.g. \cite{2016ApJ...830...51S} who use $B$, $J$, and $K$ filters). We split our $uJK$ diagram (top-left panel of Figure~\ref{fig:stars}), containing only sources with $K$-band $S/N>3$ and $K_{tot} < 25.75$, with dashed lines defined by Equation~\ref{eqn:uJK_eqn}. These dashed lines were marked where we see a clear division between the stellar sequence at the bottom and extended sources at the top. We classify sources as stars if they fall below the dashed lines. We indicate the brightness in the $K$-band of so-selected stars through purple-yellow shading; the distribution of these stars is then shown in the bottom-right panel of Figure~\ref{fig:stars} with the same color shading.


\begin{equation}
\begin{split}
    (J-K) < 0.1727 (u-J) - 0.6546 \;\; and \\
    (J-K) < 0.0176 (u-J) - 0.1112
\end{split}
\label{eqn:uJK_eqn}
\end{equation}

Then, to complement the star selection described above at the bright end, we again considered sources with $K$-band $S/N>3$ and plotted them in the $\log{(\fluxrad)}$ vs $K_{tot}$ diagram. This method of selecting stars works well for bright sources, as it identifies point-like bright objects, i.e., most likely stars \citep{2018ApJS..235...14S} with a clear separation at $K_{tot} \lesssim 20$ AB as shown in the top-right panel of Figure \ref{fig:stars}. Here, the \fluxrad\ is the half-light radius provided by \SE\, i.e., the radius at which 50\% of the \fluxauto\ is enclosed. The stars are selected as falling below the dashed lines as defined by Eq.~\ref{eqn:fluxrad_v_K} and having $K_{tot}<20$ AB. As can be seen, the stars form a clear sequence shown in blue; their $K_{tot}$ distribution is further plotted in the bottom-right panel of Figure \ref{fig:stars}.

\begin{equation}
\begin{split}
    \log{(\fluxrad)} < -0.17~K_{tot} - 2.1 \;\; and \\
    \log{(\fluxrad)} < -0.005~K_{tot} - 0.205
\end{split}
\label{eqn:fluxrad_v_K}
\end{equation}

We finally combined the stars selected from both methods and flagged them as \starflag\ = 1 in our photometric catalog, and plotted (in black) their fraction in the overall catalog as a function of $K_{tot}$ and their $K_{tot}$ distribution in the bottom-left and bottom-right panels of Figure \ref{fig:stars}, respectively. The two methods of selecting stars using the $u-J$ versus  $J-K$ color-color diagram and the $\log{(\fluxrad)}$ vs $K_{tot}$ diagram agree well where they overlap at $K_{tot} < 20$ AB. This can be seen by the fact that the black histogram (showing all stars; \starflag =1) almost perfectly overlaps the boundary of the purple-yellow histogram (showing uJK-selected stars), meaning that the blue histogram (showing the $\log{(\fluxrad)}$ vs $K_{tot}$-selected stars) do not introduce too many additional stars except at $19 < K_{tot}$ [AB] $< 20$.

\subsection{Masking bright stars and other artifacts}\label{subsec:masking}

In general, the photometry of the galaxies can be contaminated if they lie close enough to artifacts created by the foreground bright Milky Way stars. Some of the artifacts that bright stars can introduce, depending on the telescope design, are diffraction spikes, blooming (leakage of the flux along the horizontal axis of the star), halos, ghosts, and others (see, e.g., \citealp{2019PASJ...71..114A, 2022PASJ...74..247A} for a detailed explanation of such artifacts). The size and extent of these features are mostly predictable as they usually depend on the brightness of the stars. Therefore, we created masks based on brightness-based empirical relationships for such artifacts, and hand-drew masking regions when needed.

For UDS DR11 bands ($J$, $H$, and $K$), we derive empirical relationships of the sizes for circular patches to mask halos around bright stars and rectangular patches to mask diffraction spikes around bright stars. As such features only become visible for stars brighter than K$_{tot} <$ 16 AB, we do not apply these masks for dimmer stars. Moreover, in the UDS DR11 bands, stars also produce ghost patterns which are very faint, small, and roughly circular residual repeating patterns along the vertical and horizontal axes of the stars. For this, a mask was created for the DR8 data release by \cite{2008ApJ...685L...1Q} which we adapted and extended to the footprint of the UDS DR11 data images. Some other artifacts, e.g., cosmic rays, random streaks generated by the detectors in the UDS DR11 images, or image borders with no data or severely contaminated data, do not behave predictably. For such artifacts, we drew the masking regions manually. It should also be noted that any area masked in the $K$-band image was also masked in all of the other bands as we only consider the photometry of the $K$-band detected and unmasked objects. The objects detected in the K-band image and masked are denoted by the flag \detcontam\ = 1 in our photometric catalog, and \detcontam\ = 0 otherwise.

For the HSC bands ($g$, $r$, $i$, $z$, $y$, $NB0816$, and $NB0921$), we started from the list of bright objects (mag $<$ 18 AB in each band), identified for masking, provided with the third public data release - PDR3 \citep{2022PASJ...74..247A}. However, we did not use the corresponding masks as they seem to be too aggressive at least within the tract 8523 applicable for our catalog, and in applying these masks, we would lose a lot of valuable data. Therefore, we self-determine the empirical relationships of brightness vs. size of the halo around bright objects, and brightness vs. length/width of the rectangles for horizontal blooming around bright objects. We masked circular patches around all bright objects with mag $<$ 18 AB for each band as in \cite{2022PASJ...74..247A}, and masked horizontal rectangular patches around objects brighter than 14.2 AB, where blooming becomes visible. Furthermore, only for the HSC-$y$ band, do we see visible vertical diffraction spikes for objects brighter than 14.2 AB, which we masked using rectangular patches whose sizes were again determined empirically.

For VIDEO-$Y$ and VIDEO-$z$, to mask the halos around bright objects, we utilized the brightness vs. radius relationships of HSC-$y$ and HSC-$z$, respectively. As these VIDEO and HSC bands have similar wavelengths, the same objects will have similar intrinsic brightness between them. However, the PSF changes between the HSC bands and VIDEO bands so the brightness vs. radius relationships of HSC-$y$ and HSC-$z$ will not be applicable directly. We find that scaling them down by a factor of $\sim$ 0.5 works well for VIDEO-$Y$ and VIDEO-$z$. On the other hand, for VIDEO-$H$ and VIDEO-$K_{s}$, we derived their empirical relationships to mask halos around bright stars for objects brighter than the total magnitudes of 16 AB in both bands. Besides, artifacts such as diffraction spikes, ghosts, etc. within VIDEO-$zYHK_{s}$ bands do not appear predictable, and therefore, we mask such appearances manually.

For MegaCam$-uS$, we once again masked halos around bright stars using an empirically determined relationship for stars with a total magnitude less than 19  AB, and append them with hand-drawn masks of the bad regions that could not be easily predicted.

\subsection{\textit{K}-band Completeness} \label{subsec:completeness}

\begin{figure}[htb!]
\includegraphics[width=\columnwidth]
{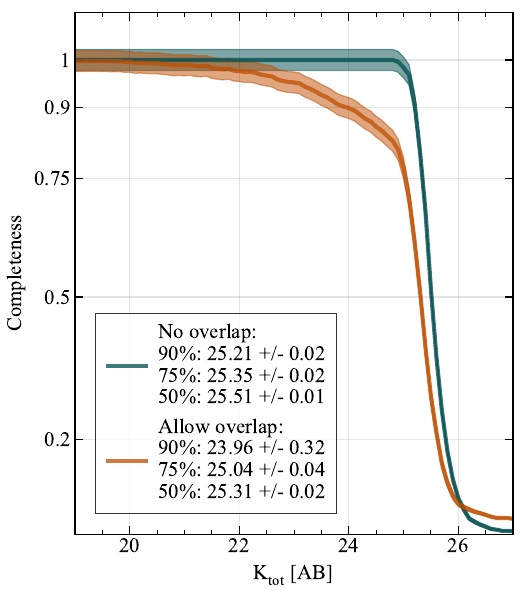}
\caption{$K$-band detection completeness as a function of $K$-band total magnitude derived from the recovery of injected stars both allowing for overlap (blending) in orange and not allowing for overlap in teal. The $K$-band magnitudes corresponding to the 50\%, 75\%, and 90\% detection completeness are also indicated for both cases.}
\label{fig:completeness}
\end{figure}

The completeness of our catalog is limited by the sensitivity in the $K$-band, the band used to perform the detection of sources, and the deblending capabilities of \SE\ with our chosen parameter settings. We estimated the detection completeness for point sources as a function of $K$-band magnitude by calculating the recovery fraction of injected simulated point sources into the $K$-band image. To simulate the point sources, we selected a bright, unsaturated, and isolated star randomly from our $K$-band image and scaled its flux up or down before injecting 2000 copies of it at random locations in the image. We repeated this for the magnitude range $19<K_{tot}<28$ with a step of 0.1 mag. Then, we ran \SE\ with the exact same settings as used for our catalog detection and as shown in Table~\ref{tab:SEparams}, and calculated the recovery fraction as a function of $K$-band magnitude. We performed this test in two different ways: 1) not allowing the injected sources to fall on previously detected sources with the help of the segmentation map, to avoid blending of objects and maximize the recovery fraction (teal curve in Fig.~\ref{fig:completeness}); 2) allowing for the injected sources to fall on the image completely randomly, hence allowing for the overlap with previously detected sources to estimate a more realistic recovery fraction (dark orange curve in Fig.~\ref{fig:completeness}).

The 90\% detection completeness of our catalog, when no overlap is allowed, is $K_{tot}$ = 25.21 +/- 0.02 AB. On the other hand, a more conservative estimate for the detection completeness, when the overlap of sources is allowed, is found to be $K_{tot}$ = 23.96 +/- 0.32 AB, as shown in Figure 5. These errors (and those shown in Fig. 5) are the Poisson errors from counting the recovered number of injected stars. Throughout the rest of the paper, we will adopt the upper 1-$\sigma$ envelop of the completeness curves shown in Figure 5, specifically $K_{tot}$ = 25.2 AB (no overlap allowed) and $K_{tot}$ = 24.3 AB (allowing overlap) for the 90\% detection completeness.

\begin{figure}[hbt!]
\includegraphics[width=\columnwidth]
{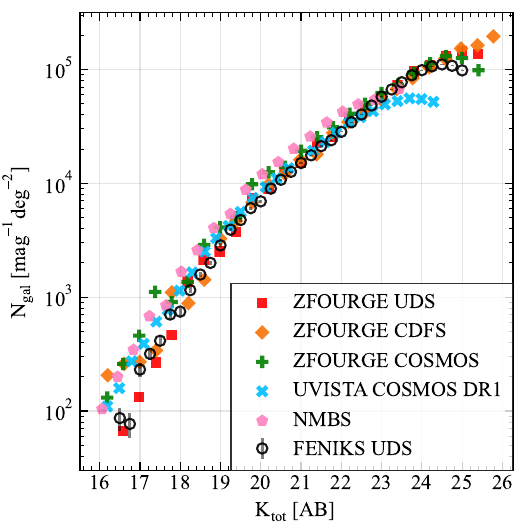}
\caption{Completeness-corrected $K$-band number counts of galaxies in our catalog (black empty circles) plotted as a function of $K$-band total magnitude, and compared with similar measurements from the literature. For our catalog, we only use galaxies with reliable photometry (\usephot\ = 1). The adopted completeness correction is estimated using the detection completeness derived allowing for the overlap of sources (orange curve in Fig.~\ref{fig:completeness}). Furthermore, we omit plotting the number counts for magnitudes corresponding to less than $\sim 70\%$ completeness.}
\label{fig:Ncounts}
\end{figure}

The completeness-corrected galaxy number counts as a function of $K$-band magnitude are shown in Figure~\ref{fig:Ncounts}, along with the number counts from other similar surveys, finding general good agreement. Here, we corrected our galaxy number counts using the detection completeness curve evaluated while allowing for the overlap of sources, excluding fainter magnitudes with less than $\sim 70\%$ completeness.

\subsection{Catalog Format/Typical selection for galaxies} \label{subsec:catformat}

We present the various quantities associated with each parameter included in the photometric catalog in detail in Table~\ref{tab:cat}. To select a robust and complete sample of galaxies with reliable photometry for an extragalactic study where completeness is important, we recommend at least applying the following selection cuts: \usephot\ $=$ 1, $0.2<z<8.0$, $\log{(M_{*}/M_{\odot})}>8$, and $K_{tot}<24.3$ (the 90\% completeness level from the upper 1-sigma envelop of the completeness curve estimated when allowing for overlap - the upper envelope of the dark orange curve in Fig.~\ref{fig:completeness}). This results in $136, 235$ sources. Using a more relaxed estimate of the completeness - $K_{tot}<25.2$ (the 90\% completeness level from the upper 1-sigma envelop of the completeness curve estimated when no overlap is allowed - the upper envelope of the teal curve in  Fig.~\ref{fig:completeness}), while keeping all other cuts the same gives $188, 564$ sources.

\section{Redshifts and Stellar Population Parameters}\label{sec:z-stellar}

\subsection{Photometric Redshifts by \eazypy}\label{subsec:photoz}

We derived the photometric redshifts using the software \eazypy\ \citep{Brammer_eazy-py_2021}, a python implementation of \EAZY\ \citep{2008ApJ...686.1503B}. It uses a linear combination of spectral energy distribution (SED) templates to fit the photometric data points for catalog sources. To fit our SEDs, we chose the Flexible Stellar Population Synthesis (\texttt{FSPS}) templates of \cite{2009ApJ...699..486C} and \cite{2010ApJ...712..833C}, a set of 12 templates denoted as \texttt{tweak\_fsps\_QSF\_12\_v3} available among the template sets provided by \eazypy, designed to target the various types of stellar populations. However, any stellar population synthesis model is hard to perfectly calibrate (e.g. due to errors in stellar evolution tracks, missing spectral features, etc.). Moreover, the vast range of star formation histories and dust extinction in galaxies within a cosmic epoch, but also over time, are difficult to fully reproduce with any given set of templates \citep{2008ApJ...686.1503B}. Therefore, to mitigate such uncertainties, \eazypy\ utilizes a wavelength-dependent template error function that captures these errors. Besides, \eazypy\ allows for the use of an apparent magnitude prior as a function of redshift, which we employed on our detection UDS-$K$ band, that assigns a lower probability of finding a bright galaxy at a high redshift than at a low redshift.

\startlongtable
\begin{deluxetable*}{llc}
\tablecolumns{3}
\tablewidth{\textwidth} 
\tablecaption{Contents of the Photometric catalog \label{tab:cat}}
\tablehead{
Column & Description & {units/range}
}
\startdata 
\texttt{ID}& Object identifier& -\\ 
\texttt{X, Y}& X and Y image coordinates. Pixel scale = 0.2684\arcsec/pixel& pixel\\
\texttt{RA, DEC}& {Right Ascension and Declination}& {J2000 deg}\\
\texttt{fcol\_X, ecol\_X}& color aperture flux and error. ZP = 25~AB& -\\
\texttt{w\_X}& {``weight": relative coverage}& 0 - 1\\
\texttt{fauto\_Kuds, eauto\_Kuds}& Flux and error in the UDS $K$-band within the Kron-like elliptical aperture.\\{}& ZP = 25~AB& -\\
\texttt{ftot\_Kuds, etot\_Kuds}& Total flux and error in the UDS $K$-band. ZP = 25~AB& -\\
\texttt{auto\_to\_tot\_corr}& correction factor to multiply  fauto\_Kuds or eauto\_Kuds to get \\&the total flux or error&  $>$ 1\\
\texttt{fD3\_Kuds, eD3\_Kuds}& Flux and error in the UDS $K$-band within the aperture diameter of 3\arcsec.\\{}& ZP = 25~AB & -\\
\texttt{fD5\_Kuds, eD5\_Kuds}& Flux and error in the UDS $K$-band within the aperture diameter of 5\arcsec.\\{}& ZP = 25~AB& -\\
\texttt{Kronradius\_Kuds}& circularized Kron radius & \arcsec\\
\texttt{aper}& Diameter of the color aperture& \arcsec\\
\texttt{aper\_tot}& Diameter of the AUTO FLUX aperture, the Kron-like aperture& \arcsec\\
\texttt{aper\_to\_tot\_corr}& correction factor to multiply \texttt{fcol\_X} or \texttt{ecol\_X} to get the total flux or error& $>$ 1\\
\texttt{r50\_Kuds}& Half-light radius in the UDS $K$-band given by \SE\ &  \arcsec\ \\
\texttt{ellipticity\_Kuds}& Ellipticity in the UDS $K$-band as output by \SE\ &  - \\
\texttt{PA\_Kuds}& Position Angle (east of north) in the UDS $K$-band as output by \SE\ &  deg\\
\texttt{flags\_Kuds}&  \SE\ flags in the UDS $K$-band as output by \SE\ & -\\
\texttt{class\_star}& Star/Galaxy classifier of  \SE\ & 0 - 1\\
\texttt{star}& Binary flag for stars derived as described in \S\ref{subsec:stars};&{}\\{}&\texttt{star=0} means that the object is classified as a star& {0 or 1}\\
\texttt{gaia\_star}& Binary flag for stars classified in the Gaia catalog\footnote{\url{https://www.cosmos.esa.int/web/gaia/dr3}} with $>$ 95\% confidence;&{}\\{}&not included in the \texttt{star} flag above& {0 or 1}\\
\texttt{opt\_nir\_maxflags}&maximum of the FLAGS output by \SE\ in each band& {}\\
\texttt{det\_contam}& Binary flag indicating contamination in the detection band (UDS-$K$)& {0 or 1}\\
\texttt{use\_phot}& Binary flag indicating galaxies with reliable photometry having \texttt{star} = 0,\\{}& \texttt{det\_contam} = 0, $S/N$ (calculated using color aperture flux/error in the &{}\\{}& UDS $K$-band) $>$ 3, and \texttt{nusefilt} (parameter in the \eazypy\ output catalog&{}\\{}& quoting the number of filters with data) $>$ 7&0 or 1\\
\texttt{z\_grism}& Binary flag indicating if the corresponding \texttt{z\_spec} came from grism spectroscopy& 0 or 1\\
\texttt{z\_spec}& Spectroscopic redshift& -
\enddata
\end{deluxetable*}

Due to imperfect background subtraction, uncertainties in the PSF matching, and other effects, the zero-points (ZPs) of each PSF-matched band may need small corrections. Following \cite{2016ApJ...830...51S} and \cite{2018ApJS..235...14S}, we determined these ZP corrections iteratively within \eazypy, using only sources with non-grism spectroscopic redshifts, i.e., sources with \texttt{use\_phot}=1, \texttt{z\_spec}$>$0, and \texttt{z\_grism}=0, adjusting the individual band ZPs at each iteration to minimize the residuals between the observed photometry and the best-fit template photometry (see \citealt{2016ApJ...830...51S} for a detailed description on the ZP correction determination). The ZP corrections so determined, as multiplicative factors to scale the flux, are listed in Table~\ref{tab:filters}.

\begin{figure}[hbt!]
\includegraphics[width=\columnwidth]
{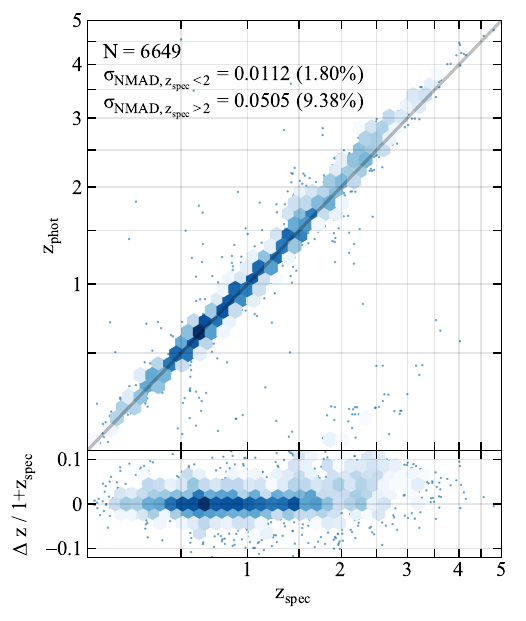}
\caption{{\it Top:} \eazypy\ derived photometric redshifts (z$_{phot}$), vs spectroscopic redshifts (z$_{spec}$) of objects with \texttt{use\_phot} = 1 and $K_{tot} < $ 25.2 AB. The $\sigma_{\textup{NMAD}}$ of objects with z$_{spec} < $ 2 and z$_{spec} > 2$ are indicated along with the corresponding outlier ($\frac{\Delta z}{1+z_{spec}}>0.15$) percentages in parenthesis. {\it Bottom:} $\Delta z / (1 +  z_{spec})$ as a function of $z_{spec}$. In both panels, the darkness of the blue hexagon bins indicates the density of the objects. If the number of objects within the hexagon bins is less than 5, the individual objects are shown as blue scatter points.}
\label{fig:zphotspec}
\end{figure}

To assess the quality of photometric redshifts, we compare them with the available spectroscopic redshifts from the following surveys: UDSz survey\footnote{\url{https://www.nottingham.ac.uk/astronomy/UDS/UDSz/}} (\citealp{2013MNRAS.433..194B, 2013MNRAS.428.1088M}; Maltby et al. in prep.), 3D-HST survey (\citealp{2016ApJS..225...27M,2012ApJS..200...13B}), SDSS-IV DR15 \citep{2017AJ....154...28B}, 6DF Galaxy Survey \citep{2009MNRAS.399..683J}, VIPERS PDR2 \citep{2018A&A...609A..84S}, PRIMUS (\citealp{2011ApJ...741....8C, 2013ApJ...767..118C}), IMACS \citep{2014ApJ...783..110K}, and others (\citealp{2013ApJS..208...24L, 2013A&A...559A..14L, 2007A&A...474..473G, 2007AJ....133..186L, 2010MNRAS.401..294S, 2005ApJ...634..861Y, 2008ApJS..176..301O, 2010MNRAS.402.1580O, 2009ApJ...701.1398S, 2017MNRAS.472..273C, 2019ApJ...877...81M, 2018MNRAS.479...25M}). All of these spectroscopic redshifts are included in the catalog, labeled as \texttt{z\_spec}. The redshifts based on grism spectroscopy from the 3D-HST survey (flag \texttt{z\_best\_s} = 2) additionally have their \texttt{z\_grism} flag set to 1  (see Table \ref{tab:cat}). As shown in Figure~\ref{fig:zphotspec}, we see excellent agreement between $z_{phot}$ and $z_{spec}$. The normalized median absolute deviation, $\sigma_{\textup{NMAD}}$ $\sim$ 1.48 x MAD of $\frac{z_{phot}-z_{spec}}{1+z_{spec}}$ ($\frac{\Delta z}{1+z_{spec}}$, hereafter), is typically quoted as a measure of the agreement between $z_{phot}$ and $z_{spec}$. For $z_{spec} < 2$, we find $\sigma_{\textup{NMAD}}$ = 0.0112, the median in $\frac{\Delta z}{1+z_{spec}}$ of 0.0014, and the catastrophic outlier percentage ($\frac{\Delta z}{1+z_{spec}}$ $>$ 0.15)  to be $\sim 1.80\%$. At high redshift, for $z_{spec} > 2$, we find $\sigma_{\textup{NMAD}}$ = 0.0505, the median in $\frac{\Delta z}{1+z_{spec}}$ of 0.0306, and the catastrophic outlier percentage ($\frac{\Delta z}{1+z_{spec}}$ $>$ 0.15)  to be $\sim 9.38\%$.

\subsection{Photometric redshift errors using close-pairs}\label{subsec:z_pair}

\subsubsection{Close-pairs method}\label{subsubsec:z_pair_method}

The distribution of spectroscopic redshifts is dominated by low-redshift (median $z_{spec} \sim 0.87$) and bright sources (median K$_{tot}$ $\sim$ 21.5). Therefore, estimating the photometric redshift errors from the comparison with spectroscopic redshifts does not provide a reliable assessment of the quality of photometric redshifts at high redshifts. 

To reliably estimate photometric redshift uncertainties over the full ranges of photometric redshifts, magnitudes, and stellar mass, we adopted the method involving galaxy pairs, first developed by  \cite{2010ApJ...725..794Q}. This method relies on the fact that the galaxies that lie in close proximity on the sky (close pairs) actually have a high probability of being physically associated, meaning that they lie at the same redshift. This is simply because galaxies are not distributed randomly in the universe, but they rather tend to cluster together. There are of course contributions to close pairs due to random projections on the sky, but they can be accounted for statistically.

\begin{figure}[hbt!]
\includegraphics[width=\columnwidth]
{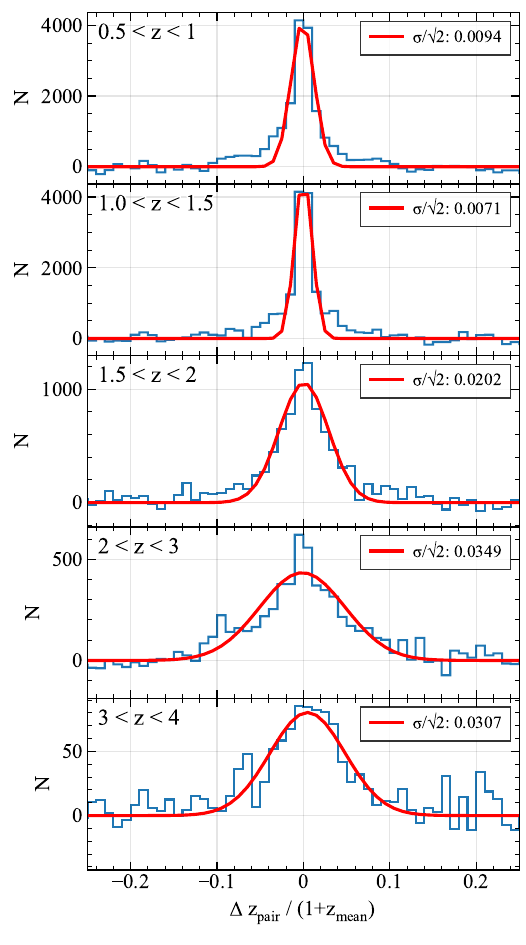}
\caption{The estimated distributions of  $\Delta z_{pair} / (1+z_{mean})$ for the true close pairs (in blue) in bins of redshifts for galaxies with $K_{tot}<23.3$ after removing contributions from the random close pairs. The fitted Gaussians are shown in red and their corresponding $\sigma$ divided by $\sqrt{2}$ values, which are the errors on the photometric redshifts, are also indicated (see \S\ref{subsec:z_pair}).} 
\label{fig:hist_zpair}
\end{figure}

In practice, we first determined all close pairs in our data catalog. If a true pair lying at the same redshift is considered, then the mean of their photometric redshifts gives a good estimate of their true redshift. Therefore, their error can then be computed by $\Delta z_{pair} / (1+z_{mean})$.  We estimated this error by fitting a Gaussian to the distribution of $\Delta z_{pair} / (1+z_{mean})$ after subtracting the contribution from random pairs. To compute the contribution to this distribution from the random projections,  we first randomized the positions in our data catalog leaving redshifts the same. Then, similar to finding pairs in the data catalog, we found the pairs in the random catalog. For example, to derive the photometric redshift errors of all the galaxies within a specified redshift interval and brighter than a given $K$-band magnitude limit, we selected all the galaxies that satisfy these criteria and their neighbors between 2.5\arcsec\ and 15\arcsec\ from the list of pairs evaluated earlier and then obtained their distribution of $\Delta z_{pair} / (1+z_{mean})$. Then we applied the same selection and methodology to the random catalog, followed by removing pairs randomly until the number of galaxies in the data catalog becomes exactly equal to the number of galaxies in the random catalog. 

Finally, we subtracted the resulting distribution of $\Delta z_{pair} / (1+z_{mean})$ of the random pairs from the original distribution which contains both random and true pairs. We repeated this process a few times to obtain an average distribution of $\Delta z_{pair} / (1+z_{mean})$ of the true pairs. The broadness of this distribution indicates the error on the photometric redshifts, with narrower distributions reflecting smaller errors and vice versa. We quantified the photometric uncertainty by fitting a Gaussian to this distribution and determining its standard deviation, $\sigma$. Since this distribution comes from pairs, we divide the $\sigma$ by $\sqrt{2}$ to obtain the average error per galaxy. We show an example of such distributions in Figure~\ref{fig:hist_zpair} in bins of redshift for a sample with $K_{tot}<23.3$. In the redshift interval, $0.5 < z < 1.0$, the average uncertainty in photometric redshifts is less than $1\%$, which progressively increases to $\sim 3.3\%$ at $3 < z < 4$.

\subsubsection{Validation of the close-pairs method}\label{subsubsec:z_pair_validation}

\begin{figure*}[hbt!]
\includegraphics[width=\textwidth]
{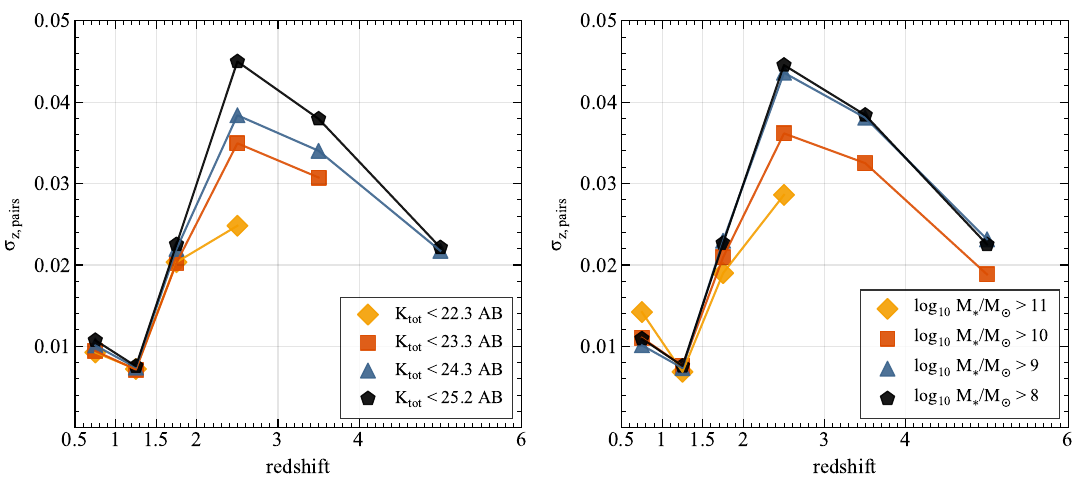}
\caption{{\it Left:} Photometric redshift errors from the close-pair analysis as a function of redshift for different $K$-band magnitude limit samples. {\it Right:} Same as the left panel but for different stellar mass limit samples.}
\label{fig:zpair_Ktot_lgM}
\end{figure*}

We also performed this analysis for different $K$-band magnitude limited samples and stellar mass threshold samples. The results of this analysis are shown in Figure~\ref{fig:zpair_Ktot_lgM}, where $\sigma_{z,  pairs}$ is the average photometric uncertainty per galaxy, equal to $\sigma/\sqrt{2}$  as described in \S \ref{subsubsec:z_pair_method} above, where $\sigma$ is the standard deviation of the Gaussian fit on the corresponding $\Delta z_{pair} / (1+z_{mean})$ distribution. The trends in the photometric redshift errors as a function of redshift are generally similar for $K$-band bright (faint) samples of galaxies and high (low) stellar mass samples, with photometric redshift errors being larger for faint and low-mass galaxies. Moreover, for both samples, the errors in $z_{phot}$ peak at $2<z<3$ across all the sub-samples, and it is found in the range of 2.5-4.5\%. This peak can be explained by the absence of the NIR medium-band filters in our catalog which, if present, would allow for a better sampling of the SEDs around the rest-frame optical Balmer/4000\AA~breaks in $z \sim 2-3$ galaxies. 

\begin{figure}[hbt!]
\includegraphics[width=\columnwidth]
{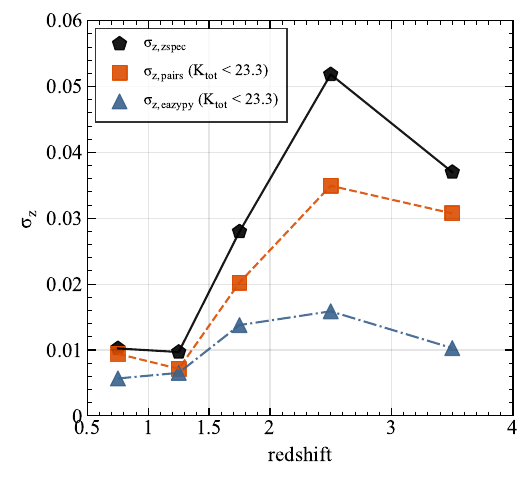}
\caption{Comparison of the $z_{phot}$ errors estimated using three different methods. The errors derived from the comparison to the spectroscopic redshifts are shown in black; the errors derived from the close-pair analysis in orange; and the median errors from the \eazypy\ redshift posterior probabilities in blue. The vertical grid lines correspond to the boundaries of redshift bins within which the close-pair errors and the \eazypy\ errors were calculated. To match the $K$-band magnitude limit of the sample with spectroscopic redshifts, the $z_{phot}$ errors using the close-pair analysis and from \eazypy\ were derived using only galaxies with $K_{tot}<23.3$.}
\label{fig:zerr}
\end{figure}

Figure~\ref{fig:zerr} shows the comparison of the photometric redshift uncertainties estimated using the three different methods, i.e., the comparison of photometric redshifts with spectroscopic redshifts where available, the close-pair analysis, and the median errors as inferred from the \eazypy\ redshift posterior probabilities. Because the sample of galaxies with spectroscopic redshifts is limited to relatively bright ($K_{tot}<23.3$) galaxies, the same $K$-band magnitude cut was adopted when deriving the photometric redshifts uncertainties with the close-pair analysis and the \eazypy\ redshift posterior probabilities. The typical uncertainty from the \eazypy\ redshift  probability distributions was derived by calculating the median of $(z_{u68}-z_{l68})/[2(1+z_{phot})]$ for all galaxies in a given redshift interval and with $K_{tot}<23.3$, where $z_{u68}$ and $z_{l68}$ are the 16-th and 84-th percentiles of the \eazypy\ redshift  probability distributions. As shown in Figure~\ref{fig:zerr}, the $z_{phot}$ uncertainties derived from the comparison with the spectroscopic redshifts are generally larger than the uncertainties derived from the close-pair analysis, whereas the $z_{phot}$ errors from the \eazypy\ redshift  probability functions are the smallest, potentially indicating a tendency by \eazypy\ to underestimate photometric redshift errors. However, we note that in the redshift interval $z<1.5$, where most of the spectroscopic redshifts are available, the $z_{phot}$ errors estimated from the three methods agree very well.

\subsection{Redshift Distribution}\label{subsec:z_dist}

\begin{figure}[hbt!]
\includegraphics[width=\columnwidth]
{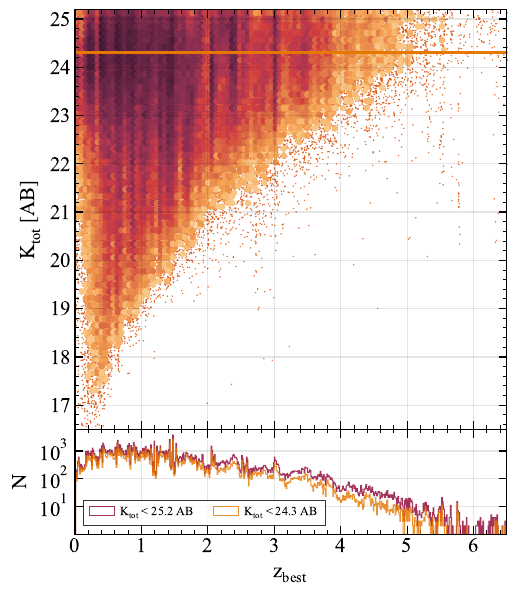}
\caption{{\it Top:} Hexagonal density plot showing the $K$-band magnitude vs. the best redshift ($z_{spec}$ if available, otherwise $z_{phot}$). Galaxies are selected with \usephot\ = 1 and $K_{tot}<25.2$. For reference, the horizontal light orange line at $K_{tot}=24.3$ indicates the adopted 90\% completeness limit} when overlapping of sources is allowed. Individual scatter points are shown where the hexagons contain less than 5 galaxies. {\it Bottom:} The redshift distribution of galaxies adopting the two different limits in $K$-band magnitude.
\label{fig:Nz}
\end{figure}

The accuracy of photometric redshifts derived by \eazypy\ allows us to capture the large-scale structure of the Universe, as clearly visible in the top panel of Figure~\ref{fig:Nz}, with the sequence of dark-colored stripes showing high-density peaks, and the light-colored stripes showing density troughs. The jaggedness of the histograms in the bottom panel of Figure~\ref{fig:Nz} also reflects the large-scale structure.

\subsection{Stellar Population Parameters}\label{subsec:stellarpop}

\subsubsection{FAST}\label{subsec:FAST}

We use the FAST code \citep{2009ApJ...700..221K} to derive  stellar population parameters such as stellar mass, star formation rate (SFR), stellar age, and dust obscuration. We adopted the stellar population synthesis library of \cite{2003MNRAS.344.1000B} and assumed a delayed exponentially declining star-formation history (SFH $\propto t~e^{-t/\tau}$), where $t$ is the time since the beginning of star formation in the range $7<\log{(t[yr])}<t_{Univ}$, with $t_{Univ}$ the age of the Universe at the redshift of the galaxy, and $\tau$ is the $e$-folding star-formation timescale in the range $7<\log{(\tau[yr])}<10$. We assumed the \cite{2003PASP..115..763C} IMF, the \cite{2000ApJ...533..682C} dust attenuation law with the visual attenuation, $A_{V}$, allowed to vary in the range $0<A_{V}$ [AB mag]$<10$, and solar stellar metallicity. We used the spectroscopic redshifts when available, otherwise, we adopted the photometric redshifts derived using \eazypy\ (see \S \ref{subsec:photoz}).

\begin{figure*}[hbt!]
\includegraphics[width=\textwidth]
{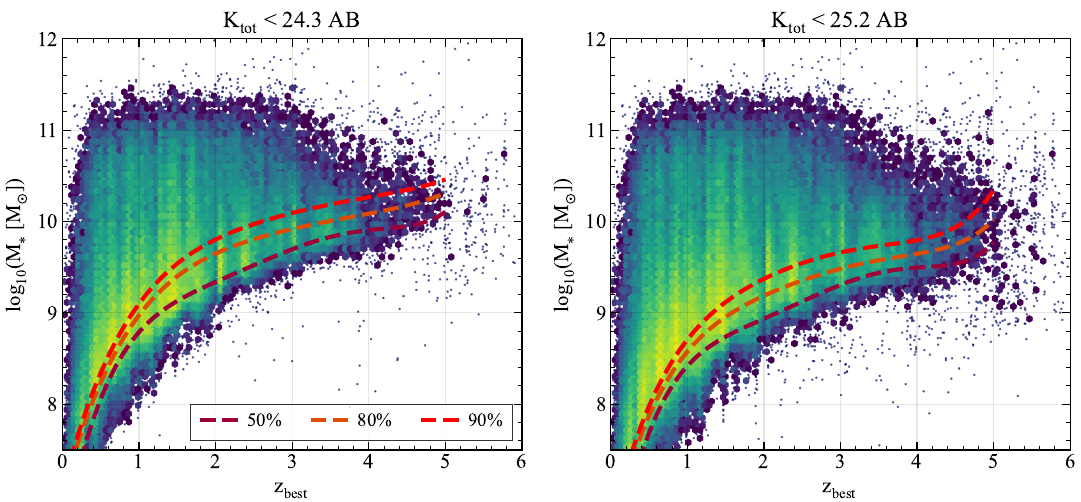}
\caption{Density maps showing the FAST-derived stellar masses as a function of the best available redshift, $z_{best}$ ($z_{spec}$ if available, otherwise $z_{phot}$) for the two $K$-band 90\% completeness limits. Bright yellow indicates high density which turns into green as the density decreases and eventually into dark blue showing the lowest densities. The left panel corresponds to the adopted 90\% K-band completeness limit of 24.3 AB derived allowing for source overlapping; the right panel corresponds to the adopted 90\% K-band completeness limit of 25.2 AB derived using the “no overlap” method} (see \S \ref{subsec:completeness} for the description on the two methods). The 50\%, 80\%, and 90\% stellar mass completeness curves as a function of redshift are also plotted as dashed brown, orange, and red curves, respectively.
\label{fig:mass_completeness}
\end{figure*}

In Figure~\ref{fig:mass_completeness}, the stellar mass of the galaxies is plotted vs the redshift, along with the 50\%, 80\%, and 90\% stellar mass completeness curves as a function of redshift. The stellar mass completeness was derived using the method described in \cite{2010A&A...523A..13P} for the two different $K$-band survey limits (90\%\ detection completeness in the $K$-band; see \S \ref{subsec:completeness}. As can be seen from Figure~\ref{fig:mass_completeness}, our catalog is complete at the 50\%(90\%) level down to $\log{(M_{\star}/M_{\odot})}$=10(10.4) at $z=5$.

\subsubsection{Dense Basis}\label{subsec:DB}

\begin{figure*}[hbt!]
\includegraphics[width=\textwidth]
{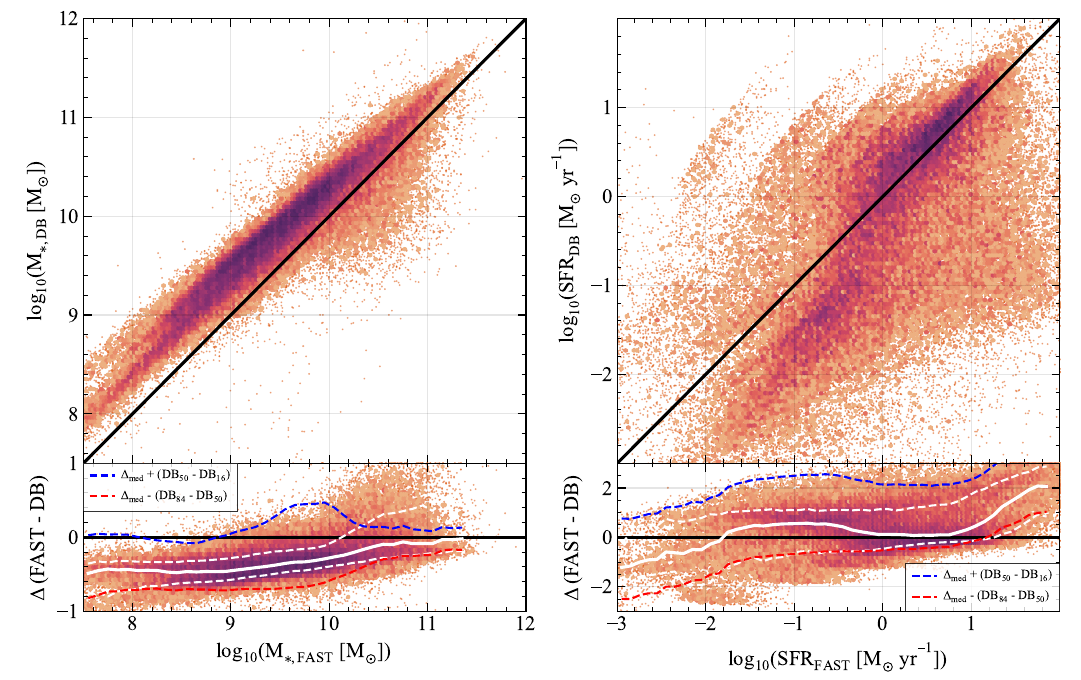}
\caption{\textbf{Left:} Comparison of stellar masses derived using FAST and Dense Basis. {\it Top:} Density map showing the median posterior stellar masses from Dense Basis as a function of best-fit stellar masses from FAST. To guide the eye, the 1-1 line is shown in black. {\it Bottom:} The offset in stellar masses between the two codes as a function of FAST stellar masses. The white solid lines show the running median calculated in FAST stellar mass bins of 0.1 dex and the white dashed lines show the corresponding 16$^{th}$/84$^{th}$ percentiles. We overplot red and blue dashed lines showing how the white solid line would shift if the 16$^{th}$ (blue) or  84$^{th}$(red) percentile Dense Basis stellar masses were used instead of the 50$^{th}$. This highlights how the uncertainties in the Dense Basis stellar masses compare to the systematic offset between the FAST and Dense Basis stellar masses. The red dashed line is derived by subtracting the running median of the difference of the 84$^{th}$ and 50$^{th}$ percentiles of the Dense Basis stellar mass posteriors. Similarly, the blue dashed line is obtained by adding the running median to the difference of the 50$^{th}$ and 16$^{th}$ percentiles of the Dense Basis stellar mass posteriors. As mentioned in the text, codes like FAST that assume parametric SFHs tend to underestimate the stellar masses as seen here. \textbf{Right:} Comparison of SFRs from FAST and Dense Basis shown in the same way as in the left panel.}
\label{fig:fast_v_db}
\end{figure*}

We also derived the stellar population properties using Dense Basis \citep{2019ApJ...879..116I}, an SED fitting code that makes use of non-parametric SFHs, which differs from codes like FAST that use only a singular function to fit the SFHs (see \S \ref{subsec:FAST}). Dense Basis is designed to extract smooth SFHs with no functional form (non-parametric SFHs) utilizing the Gaussian Process (GP; \cite{2019ApJ...879..116I}). Singular parameterizations of SFH are unable to capture different stellar populations that reside in galaxies, and therefore, are known to give biased estimates of the derived quantities from the SED fitting \citep{2022ApJ...937L..35M}. They are known to underestimate the stellar masses (\citealp{2017ApJ...838..127I, 2019ApJ...877..140L, 2020ApJ...904...33L}) as can be seen in the left panel of Figure \ref{fig:fast_v_db} which shows the comparison between the FAST-derived and Dense Basis-derived stellar masses for our catalog.

As with \eazypy,  (see \S \ref{subsec:photoz}), in Dense Basis, we use the \texttt{FSPS} templates of \cite{2009ApJ...699..486C} and \cite{2010ApJ...712..833C}, along with the nebular emission lines in \texttt{CLOUDY} (\citealp{2017RMxAA..53..385F, 2017ApJ...840...44B}). Following \cite{2023arXiv231114804C}, and based on the quality of available photometry, we constrained the shape of the non-parametric SFHs using three lookback times (\textit{t$_{25}$, t$_{50}$, t$_{75}$}) at which 25\%, 50\%, and 75\% of the stellar mass have been assembled, respectively. Furthermore, we assumed the \cite{2003PASP..115..763C} IMF, and the \cite{2000ApJ...533..682C} dust attenuation law just as we did in our FAST run described in \S \ref{subsec:FAST}. Although, unlike FAST, we imposed an exponential prior on the dust attenuation with the range $0<A_{V} $[AB mag]$<4$. Besides that, we assigned a flat prior on the specific star formation rate (sSFR=SFR/M$_{\star}$) with bounds $-14<$ sSFR [yr$^{-1}$] $<-7$, and a flat prior in log-space on metallicity within $0.01 < Z/Z_{\odot} < 2.0$. 

\subsection{Selecting Quiescent Galaxies}\label{subsec:UVJ_ugi}

\subsubsection{Using $UVJ$ diagrams}\label{subsubsec:UVJ}

\begin{figure*}[hbt!]
\includegraphics[width=\textwidth]{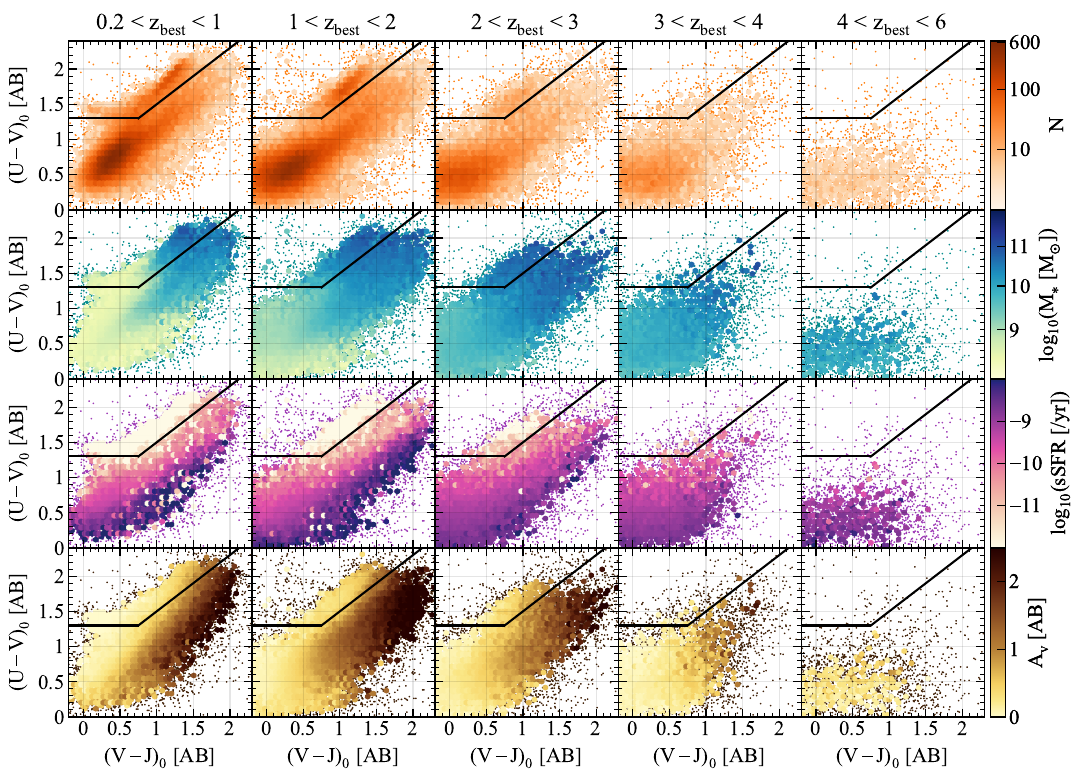}
\caption{\eazypy\ derived rest-frame $V-J$ versus $U-V$ colors of galaxies in bins of redshifts (columns), color-coded by the number of objects in the first row, followed by the FAST-derived properties - the stellar mass, the sSFR, and the dust attenuation in the $V$-band ($A_{V}$) in the second, third and the fourth rows, respectively.}
\label{fig:uvj}
\end{figure*}

The rest-frame $V-J$ versus $U-V$ color-color diagram (\textit{UVJ} diagram; hereafter) has been used to separate star-forming from quiescent galaxies, a method proposed by \cite{2009ApJ...691.1879W} and reliably utilized by many (e.g., \citealt{2011ApJ...735...86W}; \citealt{2013ApJS..206....8M}; \citealt{2016ApJ...830...51S}) at least out to $z \approx 3$ (see below for a more appropriate method for $z>3$). We derived the rest-frame magnitudes in the $U$, $V$, and $J$ filters using \eazypy\ by integrating the bandpass fluxes from the best-fitted templates of the redshifted SEDs. In Figure~\ref{fig:uvj}, we show the resulting \textit{UVJ} diagrams, each row representing a color-coding according to a different quantity - number counts, stellar mass, sSFR, and dust attenuation $A_{V}$, respectively - while the different columns correspond to six redshift intervals from $z=0.2$ to $z=6$. 

The bi-modality of star-forming and quiescent galaxies is evident in the first two panels of the first row of Figure~\ref{fig:uvj} with the star-forming galaxies mostly sitting at the bluer, lower-left section of the \textit{UVJ} diagram, and the quiescent galaxies occupying the redder, quiescent wedge at the top-left. The second row of Figure~\ref{fig:uvj} shows the trend of increasing stellar masses from the bluest corner (lower-left)  to the reddest corner (top-right), implying that more massive galaxies tend to be redder in color (either older or dustier). The ability of the of \textit{UVJ} diagram to separate star-forming from quiescent galaxies (at least out to $z\sim3$ is nicely shown by the third row of Figure~\ref{fig:uvj}, where the sSFR progressively decreases as galaxies move into the quiescent wedge of the \textit{UVJ} diagram. Finally, the bottom row of Figure~\ref{fig:uvj} shows how dust obscuration $A_{V}$ increases from the bottom-left to the top-right of the \textit{UVJ} diagram, especially outside of the quiescent galaxies' wedge. We note that the properties of the population of galaxies in the reddest corner of the \textit{UVJ} diagram are degenerate, as their red colors can be caused by their stellar population being predominantly old and quiescent, or dusty star-forming. This is also the reason why the photometric redshifts of this population of galaxies are the least reliable (e.g., \citealt{2016ApJ...830...51S}).

\subsubsection{Using $(ugi_{s})$ diagrams}\label{subsubsec:ugi}

\begin{figure*}[hbt!]
\includegraphics[width=\textwidth]{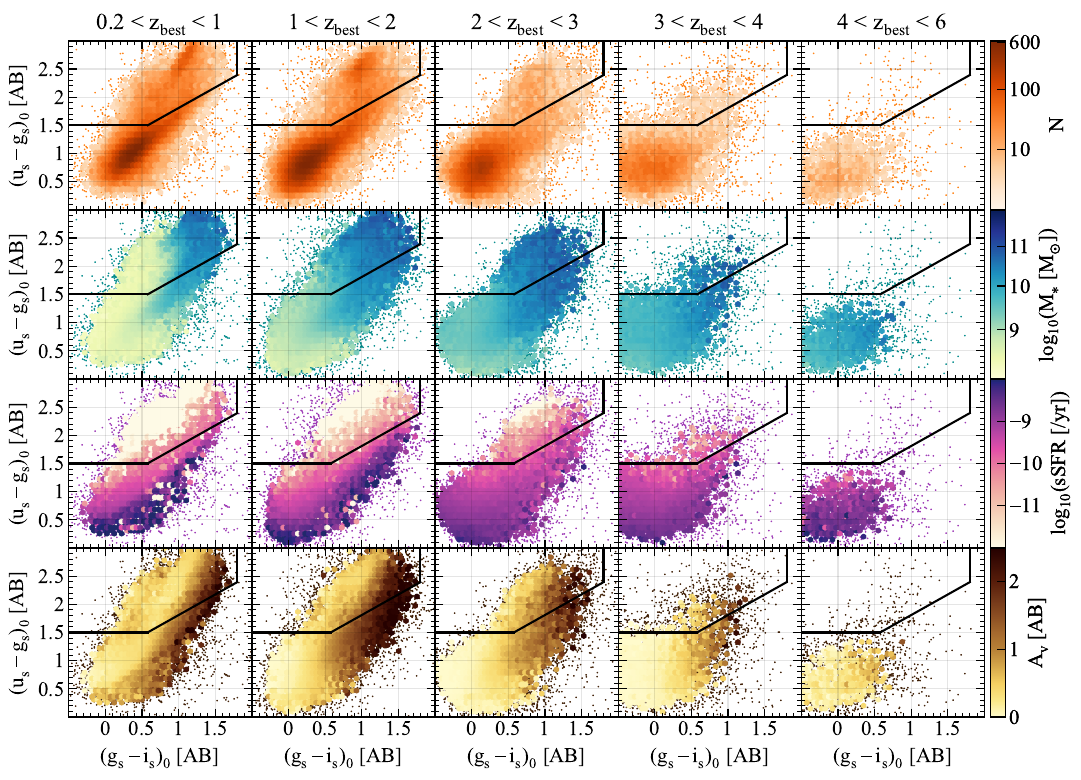}
\caption{\eazypy\ derived rest-frame $g_{s}-i_{s}$ versus $u_{s}-g_{s}$ colors of galaxies in bins of redshifts (columns). The color-coding is the same as in Fig.~\ref{fig:uvj}}
\label{fig:ugi}
\end{figure*}

Although the \textit{UVJ} diagram has been quite successful in selecting quiescent galaxies at $z \lesssim 3$, it has been known to produce contaminated samples at higher redshifts with up to $\sim$30\% galaxies undergoing considerable star-formation activity as reported by recent spectroscopic surveys (e.g., \citealt{2018A&A...618A..85S}; \citealt{2020ApJ...903...47F}). To remedy this, \cite{2023ApJ...943..166A} recently introduced a rest-frame $g_{s}-i_{s}$ versus $u_{s}-g_{s}$ color-color diagram ($(ugi)_{s}$ diagram), a selection scheme for quiescent galaxies at high redshift ($z>3$) based on the synthetic top-hat $u_{s}$, $g_{s}$, and $i_{s}$ filters centered at 2900~\AA\, 4500~\AA\, and 7500~\AA , respectively. We refer the reader to \cite{2023ApJ...943..166A} for a detailed discussion on the design of the $(ugi)_{s}$ filters. Briefly, the $u_{s}$ and  $g_{s}$ filters have narrower widths and are separated by 1600\AA\ which means that they produce a stronger $u_{s}$ -  $g_{s}$ color due to the Balmer break. Moreover, unlike the \textit{UVJ} filters, the $(ugi)_{s}$ filters avoid wavelength regions with strong emission lines for $z \gtrsim 3$ galaxies, avoiding artificially boosting the photometry and contaminating the quiescent selection. Furthermore, the rest-frame $J$ filter of the \textit{UVJ} diagram is probed, at $z \gtrsim 4$, by {\it Spitzer}-IRAC ch3 and ch4 images, which are usually not available or shallow, leading to somewhat unreliable rest-frame $V-J$ colors. The antidote to this is the rest-frame $i_{s}$-band of the $(ugi)_{s}$ diagram, which overlaps with the shorter wavelengths IRAC ch1 and ch2 at $z \gtrsim 4$, where much deeper data than from ch3 and ch4 are usually available. Therefore, this leads to more accurate $g_{s} - i_{s}$ colors, helping to break the degeneracy between dusty star-forming and quiescent galaxies. Figure~\ref{fig:ugi} shows the $(ugi)_{s}$ diagrams for our catalog similar to UVJ diagrams in Figure \ref{fig:uvj}. 

\subsubsection{Example SEDs from the catalog}\label{subsubsec:seds}

\begin{figure*}[hbt!]
\includegraphics[width=\textwidth]
{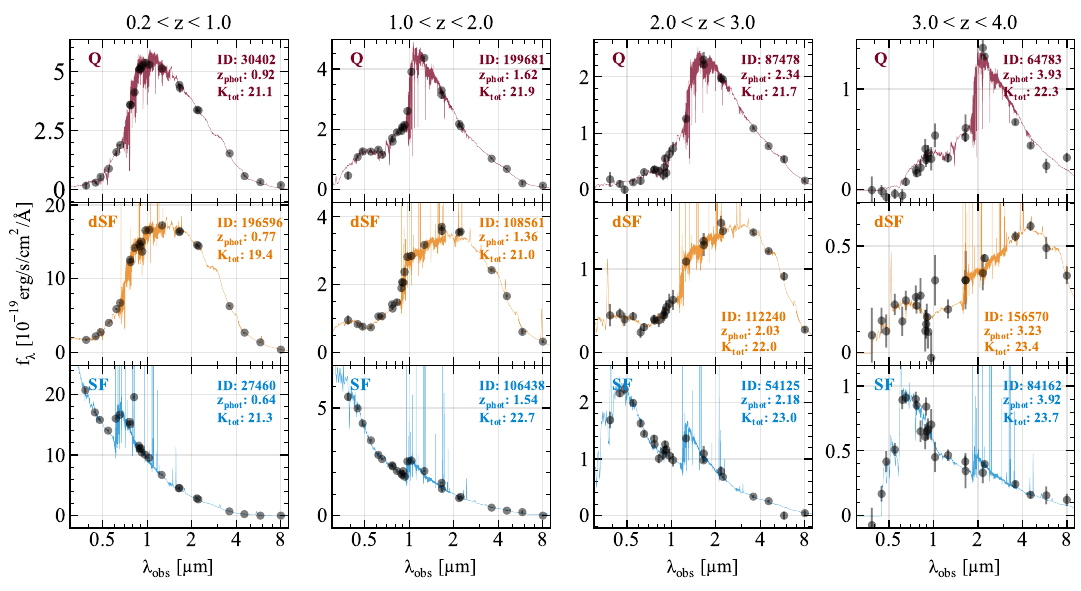}
\caption{Example SED modeling of different types of galaxies at 0.2 $<$ z $<$ 4 using \eazypy. The SED fits of quiescent galaxies (Q) are shown in dark red \textit{(top row)}, dusty star-forming galaxies (dSF) in orange \textit{(middle row)}, and the star-forming galaxies (SF) in blue \textit{(bottom row)}, as identified using the $UVJ$ and $(ugi)_{s}$ diagrams (described in \S \ref{subsec:UVJ_ugi}). To identify dSF galaxies \textit{(middle row)}, we used the wedge in the top-right corner of the $UVJ$ diagrams as defined in \cite{2022ApJ...924...25M}. Finally, the photometric data points with error bars are shown in black.}
\label{fig:seds}
\end{figure*}

Based on the $UVJ$ and $(ugi)_{s}$ diagrams, we select a handful of galaxies from our catalog and show their SEDs and modeling as output by \eazypy\ in Figure \ref{fig:seds}. The four redshift bins shown (four columns) correspond to the first four redshift bins (columns) shown in Figures \ref{fig:uvj} and \ref{fig:ugi}, and each row showcases the SEDs of quiescent (Q), dusty star-forming (dSF), and star-forming (SF) galaxies, respectively. We classify galaxies as dSF if they were found on the right side of the line represented by $U - J = 3.2$ \citep{2022ApJ...924...25M}, splitting the SF region of the $UVJ$ diagrams.

\section{Summary} \label{sec:summary}

We present a deep, PSF-matched multi-wavelength photometric catalog based on the final data release of the UDS survey (UDS DR11) covering an area of $\sim$0.9 deg$^2$. We combine ancillary data from multiple surveys: CLAUDS, SXDS, HSC, VIDEO, DeepDrill, SERVS, and SpUDS to produce a high-quality PSF-matched multi-wavelength catalog with photometry in 24 bands. The additional photometry from the $K$-split medium bands ($K_{blue}$ and $K_{red}$) of the FENIKS survey will soon be added to the catalog as soon as the data collection is completed. Many helpful flags indicating star classification, contaminated objects, etc. are also provided enabling the selection of robust stellar-mass complete samples of galaxies at $z<6$. 

In addition to the multi-wavelength photometric catalog, we also provide the catalogs of  high-fidelity photometric redshifts derived using \eazypy\ \citep{Brammer_eazy-py_2021} and a detailed assessment of the photometric redshift errors using three different methods: \eazypy\ posteriors, comparison with spectroscopic redshifts, and the z-pair analysis (see \S \ref{subsubsec:z_pair_validation}). Furthermore, the stellar population parameters were derived using FAST \citep{2009ApJ...700..221K} and Dense Basis \citep{2019ApJ...879..116I} (see \S \ref{subsec:DB} for their comparisons). All of the catalogs and the associated products such as the masks for bad regions (bright stars and other artifacts) in each band are provided publicly to enable research in the greater astronomy community.


\section{Acknowledgments}

We dedicate this paper to the memory of our dear friend and esteemed colleague Mario Nonino.

KZ acknowledges the John F. Burlingamme Graduate Fellowship in the Physics \& Astronomy Department at Tufts University for support during the latter part of this project. KZ and DM acknowledge support from the National Science Foundation under grant AST-2009442. Furthermore, CP and JAD acknowledge support from the National Science Foundation under grant AST-2009632

We extend our gratitude to the staff at the UK Infra-Red Telescope (UKIRT) for their tireless efforts in ensuring the success of the UDS project. We also wish to recognize and acknowledge the very significant cultural role and reverence that the summit of Mauna Kea has within the indigenous Hawaiian community. UKIDSS was undertaken on UKIRT while it was operated by the Joint Astronomy Centre on behalf of the Science and Technology Facilities Council of the UK, which also provided support for the Cambridge Astronomical Survey Unit (CASU) and the Edinburgh Wide Field Astronomy Unit (WFAU) that generated and served the wide field infrared public surveys from UKIRT.

Based on data products from observations made with ESO Telescopes at the La Silla Paranal Observatory as part of the VISTA Deep Extragalactic Observations (VIDEO) survey, under programme ID 179.AL2006 (PI: Jarvis).

The Hyper Suprime-Cam (HSC) collaboration includes the astronomical communities of Japan and Taiwan, and Princeton University. The HSC instrumentation and software were developed by the National Astronomical Observatory of Japan (NAOJ), the Kavli Institute for the Physics and Mathematics of the Universe (Kavli IPMU), the University of Tokyo, the High Energy Accelerator Research Organization (KEK), the Academia Sinica Institute for Astronomy and Astrophysics in Taiwan (ASIAA), and Princeton University. Funding was contributed by the FIRST program from the Japanese Cabinet Office, the Ministry of Education, Culture, Sports, Science and Technology (MEXT), the Japan Society for the Promotion of Science (JSPS), Japan Science and Technology Agency (JST), the Toray Science Foundation, NAOJ, Kavli IPMU, KEK, ASIAA, and Princeton University. 

Based in part on data from the CFHT Large Area U-band Deep Survey (CLAUDS), which is a collaboration between astronomers from Canada, France, and China described in Sawicki et al. (2019, [MNRAS 489, 5202]).  CLAUDS data products can be accessed from https://www.clauds.net. CLAUDS is based on observations obtained with MegaPrime/ MegaCam, a joint project of CFHT and CEA/DAPNIA, at the CFHT which is operated by the National Research Council (NRC) of Canada, the Institut National des Science de l'Univers of the Centre National de la Recherche Scientifique (CNRS) of France, and the University of Hawaii. CLAUDS uses data obtained in part through the Telescope Access Program (TAP), which has been funded by the National Astronomical Observatories, Chinese Academy of Sciences, and the Special Fund for Astronomy from the Ministry of Finance of China. 

This paper makes use of software developed for the Large Synoptic Survey Telescope. We thank the LSST Project for making their code available as free software at  http://dm.lsst.org

This paper is based [in part] on data collected at the Subaru Telescope and retrieved from the HSC data archive system, which is operated by the Subaru Telescope and Astronomy Data Center (ADC) at National Astronomical Observatory of Japan. Data analysis was in part carried out with the cooperation of Center for Computational Astrophysics (CfCA), National Astronomical Observatory of Japan. The Subaru Telescope is honored and grateful for the opportunity of observing the Universe from Maunakea, which has the cultural, historical and natural significance in Hawaii. 

We acknowledge ESO Programme 180.A-0776 (PI: Almaini) which produced the publicly available spectroscopic redshifts of the UDSz survey.

This work is based on observations taken by the 3D-HST Treasury Program (HST-GO-12177 and HST-GO-12328) with the NASA/ESA Hubble Space Telescope, which is operated by the Association of Universities for Research in Astronomy, Inc., under NASA contract NAS5-26555.

Funding for the Sloan Digital Sky Survey IV has been provided by the Alfred P. Sloan Foundation, the U.S. Department of Energy Office of Science, and the Participating Institutions. SDSS acknowledges support and resources from the Center for High-Performance Computing at the University of Utah. The SDSS website is www.sdss4.org.

SDSS is managed by the Astrophysical Research Consortium for the Participating Institutions of the SDSS Collaboration including the Brazilian Participation Group, the Carnegie Institution for Science, Carnegie Mellon University, Center for Astrophysics | Harvard \& Smithsonian (CfA), the Chilean Participation Group, the French Participation Group, Instituto de Astrofísica de Canarias, The Johns Hopkins University, Kavli Institute for the Physics and Mathematics of the Universe (IPMU) / University of Tokyo, the Korean Participation Group, Lawrence Berkeley National Laboratory, Leibniz Institut für Astrophysik Potsdam (AIP), Max-Planck-Institut für Astronomie (MPIA Heidelberg), Max-Planck-Institut für Astrophysik (MPA Garching), Max-Planck-Institut für Extraterrestrische Physik (MPE), National Astronomical Observatories of China, New Mexico State University, New York University, University of Notre Dame, Observatório Nacional / MCTI, The Ohio State University, Pennsylvania State University, Shanghai Astronomical Observatory, United Kingdom Participation Group, Universidad Nacional Autónoma de México, University of Arizona, University of Colorado Boulder, University of Oxford, University of Portsmouth, University of Utah, University of Virginia, University of Washington, University of Wisconsin, Vanderbilt University, and Yale University.

Funding for PRIMUS is provided by NSF (AST-0607701, AST-0908246, AST-0908442, AST-0908354) and NASA (Spitzer-1356708, 08-ADP08-0019, NNX09AC95G).


\software{\texttt{Astropy} \citep{2022ApJ...935..167A}, 
          \texttt{Source Extractor} \citep{1996A&AS..117..393B},
          \eazypy\ \citep{Brammer_eazy-py_2021},
          \texttt{numpy} \citep{harris2020array}
          }



\appendix

\section{Comparison of FENIKS photometry with other catalogs}\label{sec:mag_offsets}

\subsection{Comparison with the ZFOURGE photometric catalog}\label{zfourge_offsets}

\begin{figure*}[hbt!]
\includegraphics[width=\textwidth]
{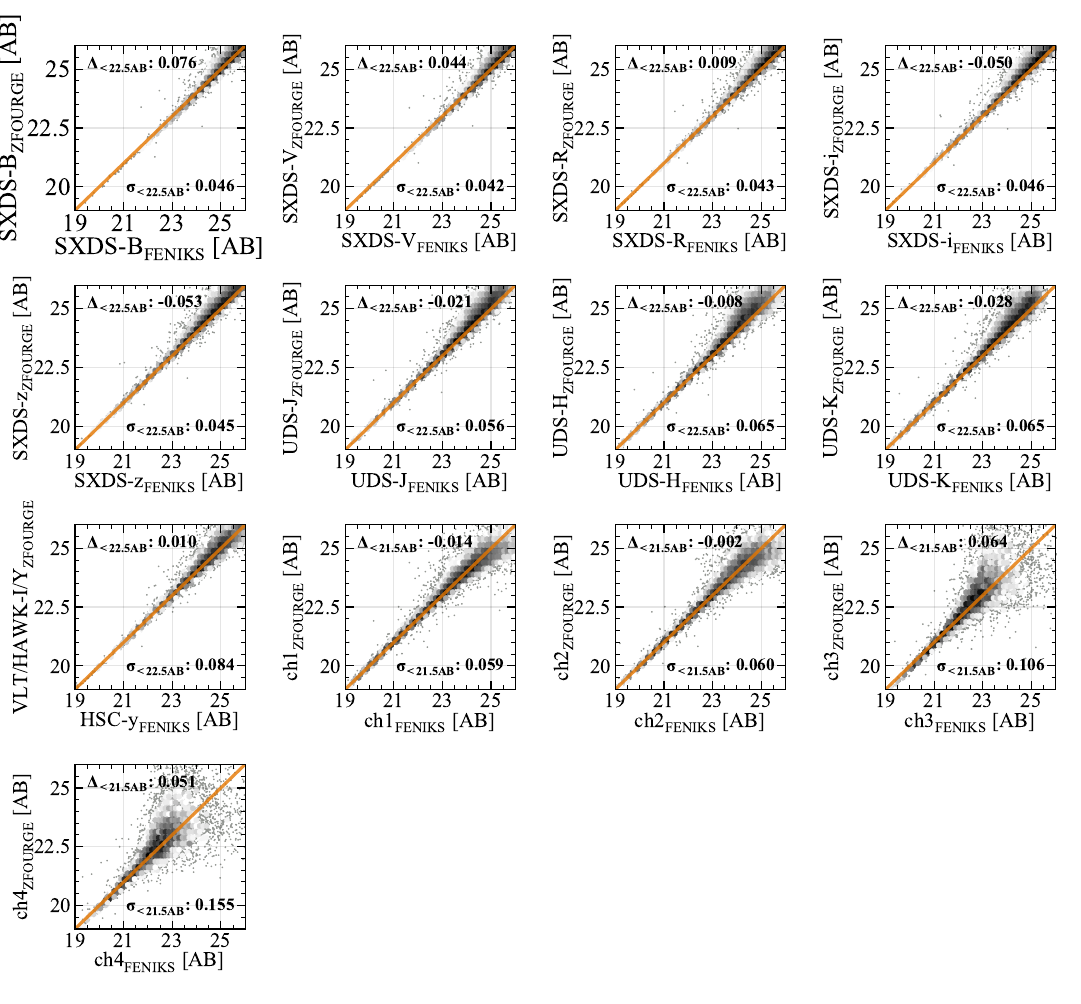}
\caption{Comparison of total magnitudes in bands common between our catalog and ZFOURGE, except for the Y-band. The ZFOURGE's Y-band photometry comes from VLT/HAWK-I \citep{2014A&A...570A..11F}. The median offsets for sources brighter than 22.5 AB mag for non-IRAC bands and 21.5 AB for IRAC bands (last four panels) are shown on the top-left of each panel. The corresponding 1-$\sigma$ values are shown on the bottom-right of each panel.}
\label{fig:zfourge_offsets}
\end{figure*}

Here we assess the quality of our photometry against the optical through mid-IR photometric catalog produced as a result of the FourStar Galaxy Evolution (ZFOURGE) survey \citep{2016ApJ...830...51S}. The ZFOURGE survey was conducted using medium-band filters of the FourStar imager \citep{2013PASP..125..654P} ranging from $1-1.8\mu m$ and consisted of three 11\arcmin x 11\arcmin pointings including one in the UDS field. We compare the total, galactic extinction corrected, and ZP-corrected photometry from our catalog with the same from the ZFOURGE catalog in Fig. \ref{fig:zfourge_offsets}.

\subsection{Comparison with the VIDEO photometric catalog (MAGAZE3NE Survey)}\label{marianna_offsets}
\begin{figure*}[hbt!]
\includegraphics[width=\textwidth]
{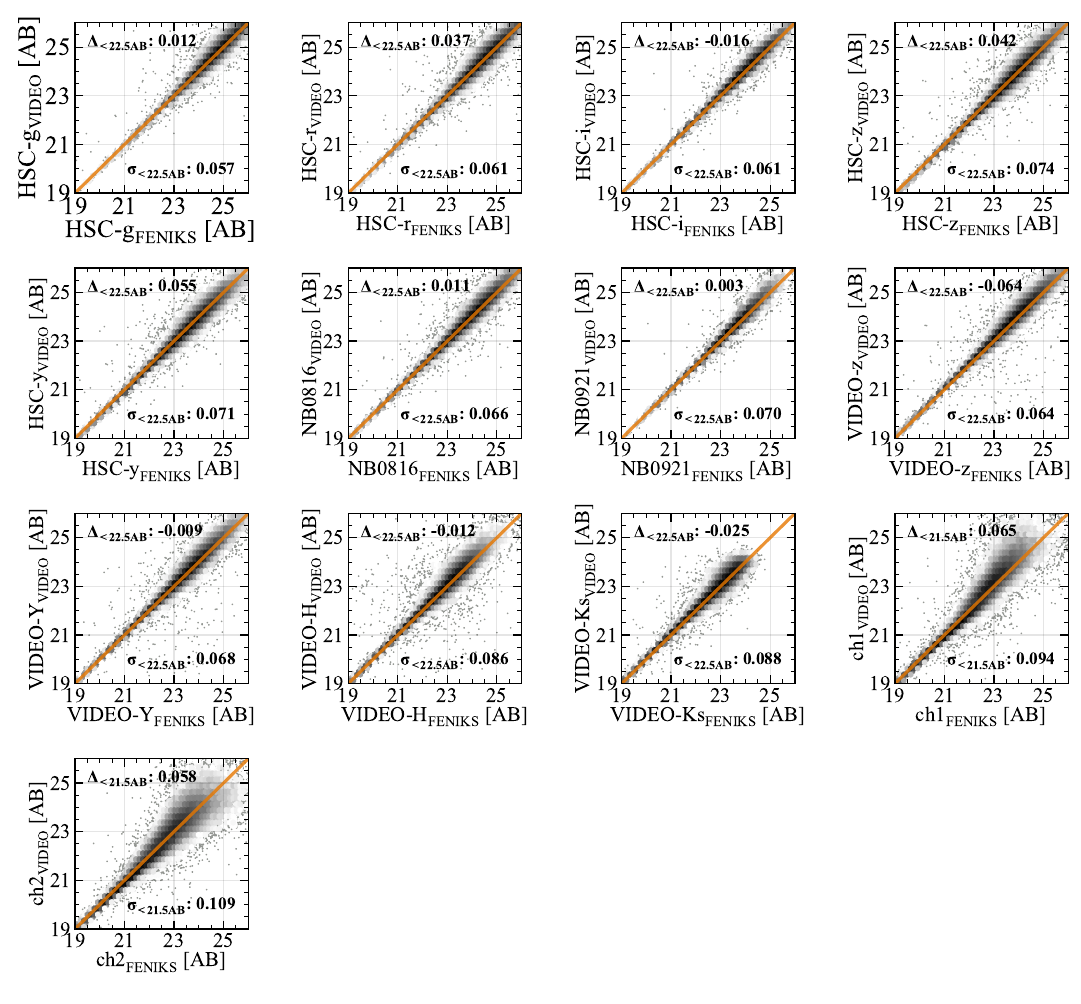}
\caption{Comparison of total magnitudes in bands common between our catalog and the VIDEO catalog constructed for the MAGAZ3NE Survey \citep{2020ApJ...903...47F} shown as in Figure \ref{fig:zfourge_offsets}.}
\label{fig:marianna_offsets}
\end{figure*}

Here we assess the quality of our photometry against the VIDEO catalogs constructed for the MAGAZ3NE survey \citep{2020ApJ...903...47F}, by Annunziatella et al. (2024; in prep.), using the same technique as described in \cite{2013ApJS..206....8M}. We compare the total, galactic extinction corrected, and ZP-corrected photometry from our catalog with theirs in Fig. \ref{fig:marianna_offsets}.

\bibliography{references}{}
\bibliographystyle{aasjournal}



\end{document}